\documentclass[aip,reprint,jcp]{revtex4-1}
\draft 
\usepackage[utf8]{inputenc}
\usepackage[T1]{fontenc}
\usepackage{lmodern}

\usepackage{graphicx}
\usepackage{amsmath}
\usepackage{amsfonts}
\usepackage{amssymb}
\usepackage{fixmath}
\usepackage{microtype}
\usepackage{hyperref}
\usepackage[all]{hypcap}
\usepackage{braket}
\usepackage{multirow}

\newcommand*{\abs}[1]{\left|#1\right|}
\newcommand*{\pvec}[1]{\vec{#1}\mkern2mu\vphantom{#1}'}
\usepackage{bm}\newcommand*{\mat}[1]{\bm{#1}}

\newcommand*{\dd}{\mathrm{d}}
\newcommand*{\ee}{\mathrm{e}}

\begin{document}

\title{Vibrationally resolved UV/Vis spectroscopy with time-dependent density functional based tight binding}

\author{Robert R\"uger}
\email{rueger@scm.com}
\affiliation{Scientific Computing \& Modelling NV, De Boelelaan 1083, 1081 HV Amsterdam, The Netherlands}
\affiliation{Department of Theoretical Chemistry, Vrije Universiteit Amsterdam, De Boelelaan 1083, 1081 HV Amsterdam, The Netherlands}
\affiliation{Wilhelm-Ostwald-Institut für Physikalische und Theoretische Chemie, Linnéstr. 2, 04103 Leipzig, Germany}

\author{Thomas Niehaus}
\affiliation{Univ Lyon, Université Claude Bernard Lyon 1, CNRS, Institut Lumière Matière, F-69622, Villeurbanne, France}

\author{Erik van Lenthe}
\affiliation{Scientific Computing \& Modelling NV, De Boelelaan 1083, 1081 HV Amsterdam, The Netherlands}

\author{Thomas Heine}
\affiliation{Wilhelm-Ostwald-Institut für Physikalische und Theoretische Chemie, Linnéstr. 2, 04103 Leipzig, Germany}

\author{Lucas Visscher}
\affiliation{Department of Theoretical Chemistry, Vrije Universiteit Amsterdam, De Boelelaan 1083, 1081 HV Amsterdam, The Netherlands}

\date{\today}

\begin{abstract}
We report a time-dependent density functional based tight-binding (TD-DFTB) scheme for the calculation of UV/Vis spectra, explicitly taking into account the excitation of nuclear vibrations via the adiabatic Hessian Franck-Condon (AH\textbar FC) method with a harmonic approximation for the nuclear wavefunction.
The theory of vibrationally resolved UV/Vis spectroscopy is first summarized from the viewpoint of TD-DFTB.
The method is benchmarked against time-dependent density functional theory (TD-DFT) calculations for strongly dipole allowed excitations in various aromatic and polar molecules.
Using the recent \texttt{3ob:freq}~parameter set of Elstner's group, very good agreement with TD-DFT calculations using local functionals was achieved.
\end{abstract}

\pacs{31.15.ee}

\maketitle

\section{Introduction}
Linear response time-dependent density functional theory (TD-DFT) based on Casida's equations~\cite{CasidaTDDFT1995} is probably the most widely used method for the simulation of processes involving excited states, such as photon absorption and emission. 
In UV/Vis spectroscopy the absorption or emission of a photon changes the electronic as well as the vibrational state of the molecule.
Thus, an accurate prediction of UV/Vis spectra needs to take the simultaneous excitation of electrons and nuclear vibrations into account.

When using TD-DFT for the calculation of absorption spectra, these vibrational contributions, which are typically much smaller than the electronic ones, often cannot be included due to the computational cost and complexity of calculating the normal modes of vibration in the electronically excited state.
The resulting absorption band is then approximated as a single line at the vertical excitation energy, which is artificially broadened through convolution with a Gaussian or Lorentzian function.
This approximation is particularly severe if the absorption spectrum is dominated by a single electronic transition, because then the structure of the spectrum is entirely determined by the vibrational fine structure.
Furthermore, the vertical excitation energy is strictly speaking not a physical observable and its use as the center of an absorption band relies heavily on error compensation.
In case of emission spectroscopy the emission of a photon occurs mostly from the lowest excited state of a given multiplicity~\cite{KashaRule1950} (Kasha's rule), so that the shape of the emission spectrum is in fact mostly determined by vibronic effects.

Based on the framework of density functional based tight-binding (DFTB)~\cite{PorezagDFTB1995,SeifertDFTB1996}, a computationally very efficient alternative to TD-DFT has been developed in the form of time-dependent DFTB~\cite{NiehausTDDFTB2001}. Here, the calculation of excited states is based on a DFTB ground state calculation and additional approximations are made to Casida's TD-DFT equations in order to avoid numerical integration at runtime. The resulting method is orders of magnitude faster than TD-DFT and has successfully been used in a wide variety of applications~\cite{tddftbapp1_doi:10.1021/jp026752s,tddftbapp2_PhysRevB.73.205312, tddftbapp3_doi:10.1021/jp071125u,tddftbapp4_doi:10.1021/ct700041v,tddftbapp5_doi:10.1021/jp065704v,tddftbapp6_10.1063/1.2715101,tddftbapp7_doi:10.1063/1.2940735,BonacicNonAdMDWithTDDFTB2009,tddftbapp9_PSSB:PSSB201100719,tddftbapp10_Fan201417}. A recent review of TD-DFTB can be found in Ref.~\citenum{NiehausTDDFTBReview2009}.

While the applicability of TD-DFT to the calculation of vibrationally resolved UV/Vis spectra has been confirmed in benchmark calculations~\cite{DierksenVibronicDFT2004,DierksenVibronicEXX2004,BaroneFrankCondon2009,BaroneFrankCondon2010,StendardoVibResDithiophene2012,BaroneVibResSpecComparison2015}, no such studies have been performed for TD-DFTB.
As such it is not clear at the moment whether the additional approximations made in TD-DFTB have a negative influence on the quality of calculated vibronic effects. 
TD-DFTB has, however, been found to yield satisfactory accuracy for both excited state geometries and excited state normal modes of small molecules~\cite{NiehausTDDFTBForces2007}, which, together with its computational efficiency, makes its application to vibrationally resolved spectroscopy promising.
In this article we investigate the applicability of TD-DFTB to the calculation of vibrationally resolved UV/Vis spectra.

This paper is structured as follows:
In Section~\ref{s:theory} we recapitulate the theory of vibrationally resolved UV/Vis spectroscopy from a TD-DFTB perspective.
We furthermore present a method to follow a particular excitation through conical intersections during an exited state geometry optimization.
In Section~\ref{s:results} we evaluate the performance of TD-DFTB for the calculation of vibronic effects for strongly dipole allowed excitations in various aromatic and polar molecules.
We compare the obtained results to both TD-DFT calculations and experimentally obtained data.
Section~\ref{s:conclusion} summarized our findings and concludes the article.

\section{\label{s:theory}Theory}

\subsection{\label{ss:tddftb}Excited states from TD-DFTB}

In the field of quantum chemistry the most commonly used density functional based approach to excited state calculations is Casida's linear response formalism~\cite{CasidaTDDFT1995}.
Here the problem of calculating excitation energies and excited states is cast into an eigenvalue equation in the space of single orbital transitions~$\hat c^\dagger_a \hat c^{\phantom\dagger}_i \ket{\psi_0}$, where~$\ket{\psi_0}$ is the Slater determinant of the occupied Kohn-Sham orbitals.
For local exchange-correlation functionals the eigenvalue problem can be written as
\begin{equation}\label{eq:CasidasEquation}
\mat \Omega \vec F = \Delta^2 \vec F \; ,
\end{equation}
where $\Delta$ is the vertical excitation energy.
For closed-shell systems the elements of the matrix~$\mat \Omega$ are given by
\begin{equation}
\Omega_{ia,jb} = \delta_{ij} \delta_{ab} \Delta_{ia}^2  + 4 \sqrt{\Delta_{ia}\Delta_{jb}} K_{ia,jb} \; .
\end{equation}
We follow the usual convention of using the indexes~$i,j$ for occupied and~$a,b$ for virtual orbitals and have abbreviated the difference in the Kohn-Sham orbital energies \mbox{$\varepsilon_a - \varepsilon_i = \Delta_{ia}$}.
The so-called coupling matrix~$\mat K$ differs between TD-DFT and TD-DFTB and also depends on the spin state of the calculated excited state.
The TD-DFTB coupling matrix is obtained from the TD-DFT coupling matrix through an approximate decomposition of the transition density into monopolar contributions and is given by~\cite{NiehausTDDFTB2001}
\begin{equation}\label{eq:TDDFTB_coupling_matrix}
K_{ia,jb} = \sum_\mathcal{AB} q_{ia,\mathcal A} \; \kappa_\mathcal{AB} \; q_{jb,\mathcal B} \; ,
\end{equation}
where the so called atomic transition charges
\begin{equation}\label{eq:AtomTransCharge}
q_{ia,\mathcal A} = \frac{1}{2} \sum_{\mu \in \mathcal A} \sum_{\nu} \Big( c_{\mu i} S_{\mu \nu} c_{\nu a} + c_{\nu i} S_{\nu \mu} c_{\mu a} \Big)
\end{equation}
are calculated through Mulliken population analysis~\cite{MullikenPopulationAnalysis1955} from the overlap and coefficient matrices~$\mat S$ and~$\mat C$.
We use capital calligraphic indexes~$\mathcal A, \mathcal B$ for atoms and Greek indexes~$\mu, \nu$ for the atomic basis functions.
The atomic coupling matrix~$\mat \kappa$ depends on the multiplicity of the calculated excitation and is given by
\begin{align}
\kappa^\mathrm{S}_\mathcal{AB} &= \gamma_\mathcal{AB} & &\text{for singlets} \\
\text{or} \qquad \kappa^\mathrm{T}_\mathcal{AB} &= \delta_\mathcal{AB} M_\mathcal{A} & &\text{for triplets},
\end{align}
where $\gamma_\mathcal{AB}$ is the normal $\gamma$-functional used in the SCC (self-consistent charge) extension of DFTB~\cite{SeifertSCCDFTB1998} and $M_\mathcal A$
is the magnetic Hubbard parameter which is also used in spin polarized ground state calculations~\cite{KohlerSDFTB2001}.

Detailed information about the excited state~$\ket{\Psi}$ can be extracted from the eigenvectors~$\vec F$ in equation~\eqref{eq:CasidasEquation}:
The electronic transition dipole moment is in TD-DFTB easily calculated as
\begin{align}
\label{eq:LROsciStrength} \braket{\psi | \hat{\vec \mu}_e | \psi_0} &= \sum_{ia} \sqrt{\frac{2\Delta_{ia}}{\Delta}} F_{ia} \, \sum_\mathcal A q_{ia,\mathcal A} \vec R_\mathcal A \; ,
\end{align}
where $\hat{\vec \mu}_e = - e \sum_i \hat{\vec r}_i$ is the electronic dipole moment operator.
Using Casida's assignment ansatz we can construct an approximate excited state wavefunction~$\ket{\psi}$ from a combination of single orbital excitations of the Kohn-Sham Slater determinant~$\ket{\psi_0}$.
\begin{equation}\label{eq:CasidaAssignmentAnsatz}
\ket{\psi} = \sum_{ia} \sqrt{\frac{2\Delta_{ia}}{\Delta}} F_{ia} \; \hat c^\dagger_a \hat c^{\phantom\dagger}_i \ket{\psi_0}
\end{equation}

Following the auxiliary functional approach\cite{FurcheTDDFTForces2002} developed by \citeauthor{FurcheTDDFTForces2002}, analytical gradients for the excitation energies have been derived by \citeauthor{NiehausTDDFTBForces2007}~\cite{NiehausTDDFTBForces2007}.
With the following definitions for several auxiliary objects
\begin{align}
U_{ia} &= \sqrt{\frac{\Delta_{ia}}{\Delta}} F_{ia} \\
\Delta q_\mathcal{A}^\mathrm{ex} &= \sum_{\mu \in \mathcal A} \sum_{\nu} P_{\mu\nu} S_{\mu\nu} \displaybreak[0]\\
U_\mathcal{A} &= \sum_{ia} U_{ia} q_{ia,\mathcal A} \displaybreak[0]\\
\Theta_\mathcal{A} &= \sum_\mathcal{B} \gamma_\mathcal{AB} \; \Delta q_\mathcal{B} \displaybreak[0]\\
\Theta^\mathrm{ex}_\mathcal{A} &= \sum_\mathcal{B} \gamma_\mathcal{AB} \; \Delta q_\mathcal{B}^\mathrm{ex} \displaybreak[0]\\
\Xi_\mathcal{A} &= \sum_\mathcal{B} \kappa_\mathcal{AB} \; U_\mathcal{B} \; ,
\end{align}
the gradient of the excitation energy~$\Delta$ can be written as
\begin{align}\label{eq:TDDFTB_forces}
\frac{\dd \Delta}{\dd \vec R_\mathcal A} = &\hspace{5.6pt} 
2 \sum_{\mathcal B \neq \mathcal A} \sum_{\substack{\mu \in \mathcal A\\\nu \in \mathcal B}} \frac{\dd H^0_{\mu\nu}}{\dd \vec R_\mathcal A} P_{\mu\nu} \\
&+ \sum_{\mathcal B \neq \mathcal A} \sum_{\substack{\mu \in \mathcal A\\\nu \in \mathcal B}} \frac{\dd S_{\mu\nu}}{\dd \vec R_\mathcal A} \left( \Theta_\mathcal{A} + \Theta_\mathcal{B} \right) P_{\mu\nu} \nonumber\displaybreak[0]\\
&+ \sum_{\mathcal B \neq \mathcal A} \frac{\dd \gamma_\mathcal{AB}}{\dd \vec R_\mathcal A} \left( \Delta q_\mathcal{A} \Delta q_\mathcal{B}^\mathrm{ex} + \Delta q_\mathcal{A}^\mathrm{ex} \Delta q_\mathcal{B} \right) \nonumber\displaybreak[0]\\
&+ \sum_{\mathcal B \neq \mathcal A} \sum_{\substack{\mu \in \mathcal A\\\nu \in \mathcal B}} \frac{\dd S_{\mu\nu}}{\dd \vec R_\mathcal A} \left( \Theta^\mathrm{ex}_\mathcal{A} + \Theta^\mathrm{ex}_\mathcal{B} \right) D_{\mu\nu} \nonumber\displaybreak[0]\\
&- \sum_{\mathcal B \neq \mathcal A} \sum_{\substack{\mu \in \mathcal A\\\nu \in \mathcal B}} \frac{\dd S_{\mu\nu}}{\dd \vec R_\mathcal A} W_{\mu\nu} \nonumber\displaybreak[0]\\
&+ 4 \sum_{\mathcal B \neq \mathcal A} \frac{\dd \kappa_\mathcal{AB}}{\dd \vec R_\mathcal A} U_\mathcal{A} U_\mathcal{B} \nonumber\\
&+ 2 \sum_{\mathcal B \neq \mathcal A} \sum_{\substack{\mu \in \mathcal A\\\nu \in \mathcal B}} \frac{\dd S_{\mu\nu}}{\dd \vec R_\mathcal A} \left( \Xi_\mathcal{A} + \Xi_\mathcal{B} \right) U_{\mu\nu} \; , \nonumber
\end{align}
where $\Delta q_\mathcal{A}$ are the Mulliken charges and $\mat D$ is the ground state's density matrix.
The exact definition of the one-particle difference density matrix~$\mat P$ and the Lagrange multipliers~$W_{\mu\nu}$ can be found in Appendix~B of Ref.~\citenum{NiehausTDDFTBForces2007}.

Analytical second derivatives have not been derived for the TD-DFTB excitation energies.
We therefore calculate the Hessian by numerical differentiation of the analytical gradient using a three point approximation with nuclear displacements of  $10^{-4}$~Bohr along the Cartesian axes.
While this increases the computational complexity of vibrational frequency calculations by a factor of~$6N_\text{atom}$, it also allows derivatives of other properties to be calculated simultaneously at no additional cost.
A useful application of this would be the calculation of the electronic transition dipole moment's gradient; a property needed for the incorporation of Herzberg-Teller effects, which are important in the correct description of weakly dipole allowed and dipole forbidden transitions.

\subsection{Vibrationally resolved spectroscopy}

Calculating absorption spectra solely from the vertical excitation energies~$\Delta$ and the corresponding electronic transition dipole moments~$\braket{\psi | \hat{\vec \mu}_e | \psi_0}$ is computationally very attractive, since the entire calculation can be performed at a fixed nuclear geometry.
This, however, completely neglects the excitation of nuclear vibrations that happens simultaneously to the electronic transition.
These nuclear effects determine the shape of an absorption band belonging to a specific electronic transition and are therefore important if an accurate absorption spectrum is required or if the spectrum is dominated by only a single band.

A more realistic description of the absorption is obtained if both electronic and nuclear effects are considered for the calculation of the transition dipole moment.
Using the Born-Oppenheimer approximation\cite{BornOppenheimer1927}
\begin{equation}
\Psi(\vec r, \vec R) = \psi(\vec r, \vec R) v(\vec R)
\end{equation}
to express the total wavefunction~$\Psi(\vec r, \vec R)$ as a product of electronic~$\psi(\vec r, \vec R)$ and nuclear wavefunction~$v(\vec R)$, the transition dipole moment can be written as
\begin{equation}
\Braket{\Psi' | \hat{\vec \mu}_e + \hat{\vec \mu}_N | \Psi} = \braket{\psi' v' | \hat{\vec \mu}_e | \psi v} + \underbrace{\braket{\psi' v' | \hat{\vec \mu}_N | \psi v}}_{=0} \; .
\end{equation}
Here~$\hat{\vec \mu}_N = e \sum_{\mathcal A} \hat{\vec R}_\mathcal A$ is the nuclear dipole moment operator.
Primed symbols denoted excited state wavefunctions and unprimed symbols the ground state wavefunctions.
The second term vanishes due to the orthogonality of the electronic wavefunctions~$\psi(\vec r, \vec R)$ and~$\psi'(\vec r, \vec R)$ for any nuclear geometry~$\vec R$.
The first term involving the electronic dipole moment is then approximated by assuming that the electronic transition dipole moment is constant in the region of configuration space where the nuclear wavefunction~$v(\vec R)$ is non-zero, which gives the Franck-Condon approximation~\cite{FranckFCF1926,CondonFCF1926,CondonFCF1928} for the transition dipole moment.
\begin{align}\label{eq:fcf_intro}
\Braket{\Psi' | \hat{\vec \mu}_e + \hat{\vec \mu}_N | \Psi} \approx \braket{v'|v} \braket{\psi'| \hat{\vec \mu}_e | \psi}
\end{align}
The square of the overlap~$\braket{v'|v}$ between the initial and final nuclear wavefunction is called the Franck-Condon factor.
For strongly dipole allowed transitions this approximation is generally sufficient.
Dipole forbidden transitions are, however, poorly described by equation~\eqref{eq:fcf_intro}, as it predicts a zero transition dipole moment.
The description of these transitions can be improved if the gradient of the electronic transition dipole moment is explicitly taken into account.
These so called Herzberg-Teller effects~\cite{HerzbergTeller1933} are beyond the scope of this article though, and we will restrict our investigation to dipole allowed transitions.

In order to calculate the overlap~$\braket{v'|v}$, some functional form for the nuclear wavefunction is required.
Within the Born-Oppenheimer approximation the nuclei move in the potential given by the electronic energy of the ground state and excited state, respectively.
Using a harmonic approximation of the Born-Oppenheimer potential around the equilibrium geometry of the respective state, the nuclear wavefunction can be approximated as a multidimensional harmonic oscillator, for which the wavefunction is well known.
\begin{equation}
v(\vec n, \vec q) = \prod_i \left( \frac{\omega_i}{\pi} \right)^\frac{1}{4} \frac{1}{\sqrt{2^{n_i} n_i!}} \; \ee^{-\frac{1}{2} \omega_i q_i^2} \; \mathcal{H}_{n_i}( \sqrt{\omega_i} q_i)
\end{equation}
Here $n_i$ is the number of energy quanta~$\omega_i$ in the $i$-th normal mode of vibration, and $q_i$ is the normal mode coordinate.
$\mathcal H_{n_i}$ is the $n_i$-th Hermite polynomial.
Before the overlap~$\braket{v'|v}$ can be calculated, one has to account for the fact that both the normal modes as well as the equilibrium geometries of ground and excited state are different.
\citeauthor{Duschinsky1937} showed~\cite{Duschinsky1937} that the normal mode coordinates of the initial and final state can be related through a linear transformation~$\pvec q = \mat J \vec q + \vec k$, which allows the overlap integral to be written as
\begin{equation}\label{eq:FranckCondonOverlapIntegral}
\braket{v'|v} = (\det \mat J)^{-\frac{1}{2}} \int v'(\pvec n, \mat J \vec q + \vec k) \, v(\vec n, \vec q) \, \dd \vec q \; .
\end{equation}
Details on the calculation of these integrals can be found in Ref.~\citenum{RuhoffRecursiveFCF1994} and~\citenum{RuhoffRatnerFranckCondonAlgs2000}.

\begin{figure}[tbp]
\includegraphics[width=\columnwidth]{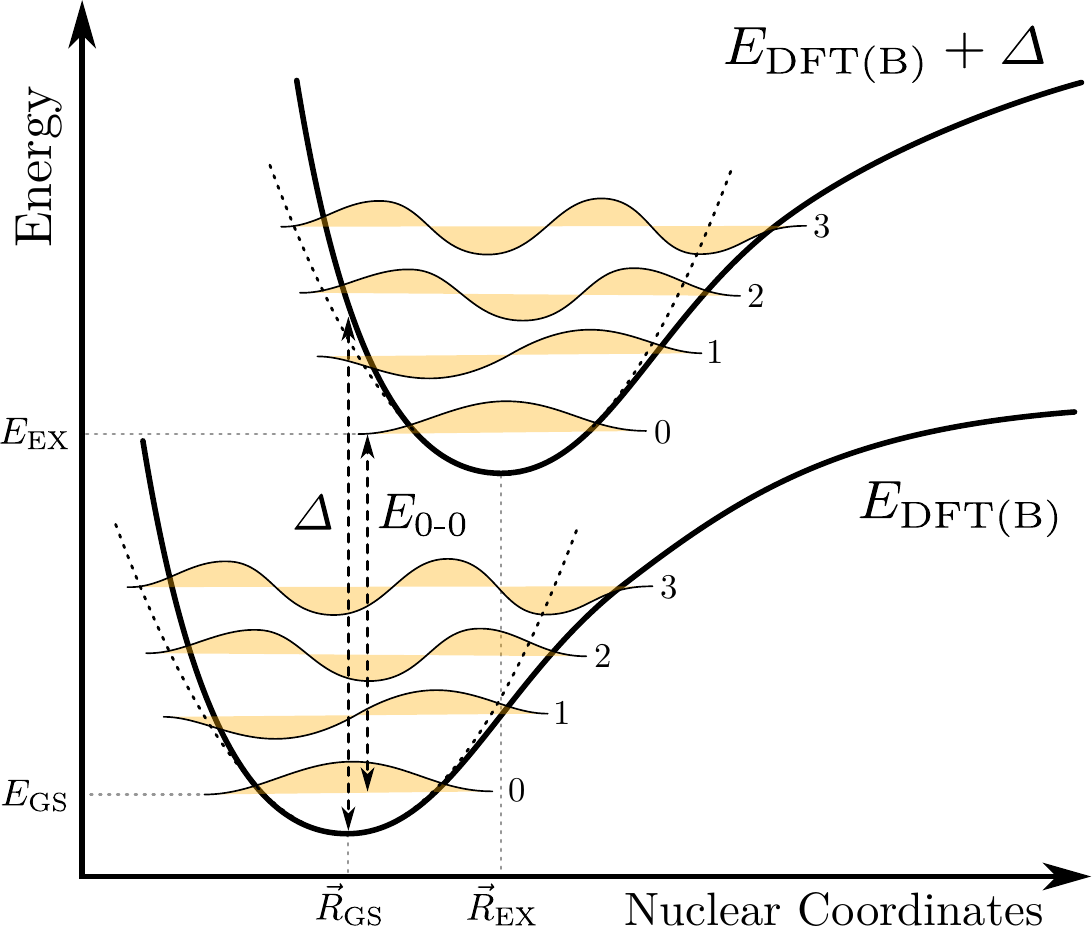}
\caption{\label{fig:fcf_illustration}Illustration of the quantities involved in vibrationally resolved spectroscopy calculations.}
\end{figure}

With the harmonic approximation for the potential experienced by the nuclei, they behave as a multidimensional harmonic oscillator for which the energy can easily be written as
\begin{equation}\label{eq:nuclear_energy}
E^\text{nuc}(\vec n) = \sum_i \omega_i \left( n_i + \frac{1}{2} \right) \; .
\end{equation}
Note that even without excitation of nuclear vibrations the nuclei in their ground state still have a vibrational zero point energy given by
\begin{equation}
E^\text{ZPE} = E^\text{nuc}(\vec 0) = \frac{1}{2} \sum_i \omega_i \; .
\end{equation}
As the thermal energy is usually small compared to the energy necessary for vibrational excitations, it is reasonable to assume that both the electrons and the nuclei are in their respective ground state when the photon is absorbed.
The total energy of the system is therefore given by the DFT(B) energy functional~$E_\text{DFT(B)}$ plus the nuclear vibrational zero point energy~$E^\text{ZPE}$, both of which are calculated at the equilibrium geometry~$\vec R_\text{GS}$ of the ground state, that is where~$E_\text{DFT(B)}$ is minimal.
\begin{equation}
E_\text{GS} = E_\text{DFT(B)} (\vec R_\text{GS}) + E^\text{ZPE}_\text{GS}
\end{equation}
After the absorption the total energy of the excited state is the sum of DFT(B) ground state energy~$E_\text{DFT(B)}$, vertical excitation energy~$\Delta$ and the nuclear vibrational energy according to equation~\eqref{eq:nuclear_energy}, where everything is evaluated at the excited state equilibrium geometry~$\vec R_\text{EX}$, that is where $E_\text{DFT(B)} + \Delta$ is minimal.
\begin{equation}
E_\text{EX}(\pvec n) = E_\text{DFT(B)} (\vec R_\text{EX}) + \Delta (\vec R_\text{EX}) + E^\text{nuc}_\text{EX}(\pvec n)
\end{equation}
The difference~$E_\text{EX}(\pvec n) - E_\text{GS}$ between ground and excited state energies together with the intensities calculated as the square of the overlap~$\braket{v'|v}$ from equation~\ref{eq:FranckCondonOverlapIntegral} determine the shape of an absorption band belonging to a specific electronic transition.
The smallest possible excitation energy called~$E_\text{0-0}$ is obtained if the nuclei remain in the ground state during the photon absorption and is given by
\begin{equation}
E_\text{0-0} = E_\text{EX}(\vec 0) - E_\text{GS} \; .
\end{equation}
An illustration of the various energies used in this article can be found in figure~\ref{fig:fcf_illustration}.

It is interesting to note that despite its widespread use for the calculation of absorption spectra the vertical excitation energy~$\Delta$ is not experimentally observable.
Its popularity stems from the fact that it is easily calculable and often a reasonable approximation for the position of an absorption band:
The vertical excitation energy~$\Delta$ is always larger than~$E_\text{0-0}$, which on the other hand is a lower bound to the photon energy at which the absorption can happen.
Depending on the details of ground and excited state potential energy surface it might well happen that these effects compensate and the vertical excitation energy~$\Delta$ is actually not too far from the band maximum.
We will later plot the vertical excitation energy into the vibrationally resolved spectra to investigate to what extent it can be used as an approximation to the absorption band's maximum.

So far we have only described absorption spectroscopy. For emission spectroscopy the only necessary modification is that the system is assumed to be in the vibrational ground state of the electronically excited state prior to the emission of the photon. All vibrational states of the electronic ground state are then valid final states. This includes the vibrational ground state, so that absorption and emission spectrum overlap at the line corresponding to the 0-0~transition.

The method outlined in this section is commonly referred to as adiabatic Hessian Franck-Condon (AH\textbar FC) since it performs an optimization of the excited state, calculates its Hessian and subsequently uses the Franck-Condon approximation for the calculation of the dipole moments.
Various other methods for the calculation of vibronic fine structures are available, a comparison of which on the basis of DFT can be found in a recent article by \citeauthor{BaroneVibResSpecComparison2015}~\cite{BaroneVibResSpecComparison2015}.
These methods generally skip one or more steps in the calculation to increase computational efficiency.
The vertical gradient Franck-Condon (VG\textbar FC) method~\cite{MacakLinearCouplingModel2000} (also called linear coupling model) for example requires neither an optimization of the excited state nor a calculation of the excited state Hessian, thereby avoiding the most costly steps in the AH\textbar FC procedure.
All of these methods can in principle also be used within the DFTB framework.
However, the purpose of the current paper is to investigate the effect of the DFTB approximations on the calculation of vibrational fine structures.
We believe that this is best done using the AH\textbar FC method, as it offers the most complete test of the DFTB framework, i.e.\ excited state geometries and vibrational frequencies as well as $E_\text{0-0}$~energies, all of which would not be tested in e.g.\ the VG\textbar FC method. We do not expect the DFTB approximations to alter the conclusions of \citeauthor{BaroneVibResSpecComparison2015}~\cite{BaroneVibResSpecComparison2015}, and have hence restricted our investigations to the AH\textbar FC method.

\subsection{\label{ss:exfollow}Following a particular excitation during geometry optimization}

\begin{figure}[tbp]
\includegraphics[width=\columnwidth]{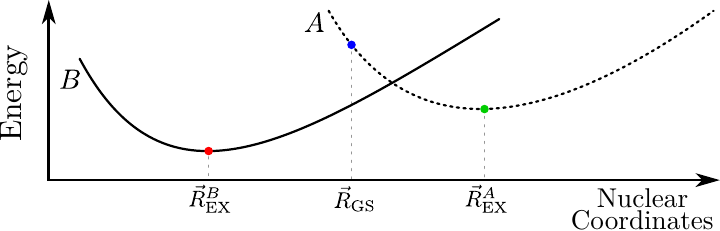}
\caption{\label{fig:exfollow}An example system where optimization on the $S_2$ potential energy surface does not converge to a minimum. Note that the ground state's potential energy surface is not shown.}
\end{figure}

In order to reliably calculate the equilibrium structure of a specific electronically excited state, we need to make sure that at every step of the geometry optimization the excitation energy gradient is calculated for the correct excited state.
If the excited states are well separated in energy, this is trivial as one can just use the gradient of the $I$-th lowest excited state, where $I$ is the number of the excited state of interest at the initial geometry.
This, however, does not work if potential energy surfaces~(PES) cross and can not only lead to finding equilibrium geometries of excited states other than the originally selected, but can also lead to completely unphysical results.
Such a case is illustrated in figure~\ref{fig:exfollow}.
Here the excited state of interest~$A$ has been identified as the $S_2$~state at the ground state's equilibrium geometry~$\vec R_\mathrm{GS}$, so that the optimization would start at the blue dot on the dotted PES.
Optimization on the PES of state~$A$ would go through the conical intersection with the other PES and end at the equilibrium geometry~$\vec R_\mathrm{EX}^A$ of state~$A$.
However, if one simply optimizes the second lowest excited state one runs into a problem at the conical intersection:
The $S_2$~surface on which the optimization would take place is the PES of state~$A$ left of the conical intersection and the PES of state~$B$ right of the intersection.
The ``minimum'' of this surface is at the intersection, but the gradient is not defined there and in practice common optimizers just oscillate around the conical intersection geometry.
In order to solve this problem generally and to reliably reach the equilibrium geometry of the originally selected excitation we need a way to follow a particular excited state through a conical intersection.

Let $\vec F_I^k$ be the eigenvector of the $I$-th excitation at the $k$-th step of the geometry optimization.
Assuming that the $I$-th excitation is the one for which the geometry is to be optimized, we would use this eigenvector to calculate the gradient which determines the nuclear geometry for the next step.
Having solved equation~\eqref{eq:CasidasEquation} at step $k+1$ of the optimization it would then be natural to look at the eigenvectors~$\vec F_J^{k+1}$ in order to check which of the new eigenvectors is most similar to~$\vec F_I^k$.
We use the following measure~$\theta_{IJ}$ for quantifying the similarity between eigenvectors in subsequent steps of the geometry optimization:
\begin{equation}\label{eq:FSimilarity}
\theta_{IJ} = \theta(\vec F_I^k, \vec F_J^{k+1}) = \abs{ \sum_{\mu\nu} F_{\mu\nu,I}^k F_{\mu\nu,J}^{k+1} }
\end{equation}
Here
\begin{equation}
F^k_{\mu\nu,I} = \sum_{ia} c^k_{\mu i} F^k_{ia,I} c^k_{\nu a}
\end{equation}
are the elements of Casida's eigenvectors in atomic orbital basis.
The similarity measure~$\theta_{IJ}$ is calculated for all~$J$ and the gradient of the most similar excitation is used to proceed with the next step of the geometry optimization.
Switching to atomic orbital basis is necessary as the molecular orbitals can vary drastically from step to step (i.e.\ their sign is undefined and their energetic ordering can change), which makes direct comparison of eigenvector elements~$F_{ia}$ difficult.

A more rigorous way to compare two excited states would be to use Casida's assignment ansatz from equation~$\eqref{eq:CasidaAssignmentAnsatz}$ to calculate the overlap~$\braket{ \psi^k_I | \psi^{k+1}_J }$.
This would also take the change in the positions of the atomic basis functions into account, which is neglected in equation~\eqref{eq:FSimilarity}.
We have, however, found that this is a small effect for the nuclear displacements typically seen from one step in a geometry optimization to the next, and that the similarity measure~$\theta_{IJ}$ very reliably selects the correct excitation, while having the advantage that it is much easier to calculate than the more rigorous overlap~$\braket{ \psi^k_I | \psi^{k+1}_J }$ between wavefunctions.

It is important to note that the procedure introduced in this section is only a technical tool that avoids certain problems during geometry optimizations of excited states. By staying on the original PES the result is also easier to interpret for the user. However, great care must be taken in situations where potential energy surfaces cross, as it can in reality happen that the system switches to another PES at the conical intersection. If this is the case, the user should manually restart the optimization from the intersection geometry on the other PES.

\subsection{Technical details}

The first step in the calculation is to perform a geometry optimization of the ground state using DFT(B).
This determines the ground state's equilibrium geometry~$\vec R_\text{GS}$ at which the vertical excitations are calculated with TD-DFT(B) and the excitation of interest identified.
The ground state geometry~$\vec R_\text{GS}$ is then slightly distorted by adding small random vectors to all nuclear positions.
This distorted geometry is then used as the initial geometry for a geometry optimization of the desired excited state, which yields the excited state's equilibrium geometry~$\vec R_\text{EX}$.
The random displacement is necessary in order to allow the excited state to break possible symmetries of the ground state, as it might otherwise well happen that the geometry optimization starts on a saddle point or local maximum, making it difficult to define an initial direction for the optimization by following the gradient.
Excitation following as introduced in subsection~\ref{ss:exfollow} is used during the geometry optimization of the excited state.
Normal modes of the ground and excited state are calculated at the respective equilibrium geometries and used to calculate the Franck-Condon factors and the vibrationally resolved spectrum.
The nuclear system is assumed to be in the vibrational ground state prior to the absorption of the photon.
The theoretical spectra are convoluted with Gaussian or Lorentzian functions of suitable widths in order to match the resolution of the respective experimental spectra.
Intensities in the theoretical spectra naturally integrate to one over the entire spectral width.
Experimental spectra were normalized in the same way, if possible.
In cases where this was impossible due to a cut off 0-0~peak or a limited spectral range, the experimental spectrum was scaled to match the intensity of a prominent line to the DFT result.
In order to facilitate a comparison of spectral shapes even when absolute excitation energies differ, all spectra have had their $E_\text{0-0}$~energy shifted to zero.

All simulations were performed with the 2016 version of the ADF modeling suite.
DFT calculations were performed with a TZP basis set and the PBE functional~\cite{PerdewBurkeErnzerhofPBEXCFunc1996}.
For the DFTB calculations we used the DFTB3 Hamiltonian~\cite{ElstnerDFTB32011} and the \texttt{3ob:freq} parameter set~\cite{Elstner3obParameters2013}.
Conceptually, using the DFTB3 Hamiltonian in the ground state calculation is slightly inconsistent, as TD-DFTB is based on the linear response of SCC-DFTB, not DFTB3.
While TD-DFTB has been adapted to the DFTB3 framework~\cite{NishimotoTDDFTB32015}, the difference in the results is negligible in practice and we used TD-DFTB in its original formulation~\cite{NiehausTDDFTB2001}.

\section{\label{s:results}Results}

In order to evaluate the loss in accuracy introduced through the DFTB approximations, we have calculated vibrationally resolved spectra with both DFT and DFTB.
In addition to the comparison between the two theoretical methods, we also compare both of them to experimentally obtained spectra.
This allows us to determine whether a deviation is indeed caused by the DFTB approximations or already present at the DFT/GGA level.


A summary of all energies and oscillator strengths can be found in table~\ref{tab:comparison}.
\begin{table*}
\begin{tabular}{c|c|c|c|c|c|c|c|c}
\multirow{2}{*}{Molecule} & \multirow{2}{*}{Excited State} & \multicolumn{3}{|c|}{DFT/PBE} & \multicolumn{3}{|c|}{DFTB} & Exp. \\\cline{3-9}  &  & $\Delta (\vec R_\text{GS})$ & $E_\text{0-0}$ & $f$ & $\Delta (\vec R_\text{GS})$ & $E_\text{0-0}$ & $f$ & $E_\text{0-0}$ \\\hline\hline
Anthracene & $1^1B_{2u}$, $L_a$ & 2.91 & 2.67 & 0.037 & 2.98 & 2.65 & 0.048 & 3.43 \\\hline
Pentacene & $1^1B_{2u}$, $L_a$ & 1.60 & 1.48 & 0.059 & 1.80 & 1.61 & 0.033 & 2.12$^\dagger$ \\\hline
Pyrene & $1^1B_{3u}$, $L_a$ & 3.38 & 3.17 & 0.211 & 3.25 & 2.97 & 0.201 & 3.84 \\\hline
Pentarylene & $1^1B_{3u}$, $L_a$ & 1.36 & 1.31 & 1.48 & 1.49 & 1.38 & 1.58 & 1.66$^\dagger$ \\\hline
Octatetraene & $1^1B_{u}$ & 3.78 & 3.53 & 1.40 & 3.83 & 3.44 & 1.14 & 4.40 \\\hline
\textit{trans}-Stilbene & $1^1B_{u}$ & 3.60 & 3.32 & 0.85 & 3.74 & 3.34 & 0.83 & 3.80$^\dagger$ \\\hline
Anisole & $1^1A'$ & 4.72 & 4.38 & 0.030 & 4.69 & 4.22 & 0.036 & 4.51 \\\hline
C480 & $1^1A$ & 3.06 & 2.73 & 0.23 & 3.19 & 2.77 & 0.22 & $\approx$ 3.22$^\dagger$ \\\hline
Bithiophene & $1^1B_{u}$ & 3.68 & 3.38 & 0.40 & 3.67 & 3.30 & 0.40 & 3.86 \\\hline
Triazoline & $1^1B_2$ & 1.67 & 1.60 & 0.0006 & 1.83 & 1.60 & 0 & 2.15$^\dagger$ \\\hline
\end{tabular}\\[5pt]
{\footnotesize $^\dagger$Measured in solution or matrix conditions. A solvent-induced shift of about\\0.1--0.2 eV should be considered when comparing with theoretical results.}
\caption{\label{tab:comparison}Summary of vertical excitation energies, $E_\text{0-0}$ energies and oscillator strengths for all example transitions from section~\ref{s:results}. All energies are given in eV. Experimental~$E_\text{0-0}$ energies from Ref.~\citenum{ExpAnthracenePyrene,ExpPentacene,ExpPentarylene,ExpOctatetraene,ExpStilbene,AnisoleExp,BaroneVibResSpecComparison2015,ExpDithiophene2,ExpTriazoline}.}
\end{table*}
For planar molecules with $D_{2h}$ symmetry, the $xy$-plane was chosen to be the plane of the molecule, with the $x$-axis along the longer axis of the molecule.

\subsection{Anthracene}

Our first example is the $\mathrm S_0 \rightarrow \mathrm S_1$ excitation in anthracene, a typical example of a dipole-allowed $\pi \rightarrow \pi^*$ transition in a polycyclic aromatic hydrocarbon (PAH).
Anthracene is a precursor to anthraquinone from which many technically important dyes such as alizarin are derived.
Furthermore, the optical properties of anthracene have recently also been discussed in the context of absorption lines of interstellar molecular clouds of which anthracene is one of the most complex constituents~\cite{AstroAnthracene}.

\begin{figure}[tbp]
\includegraphics[width=\columnwidth]{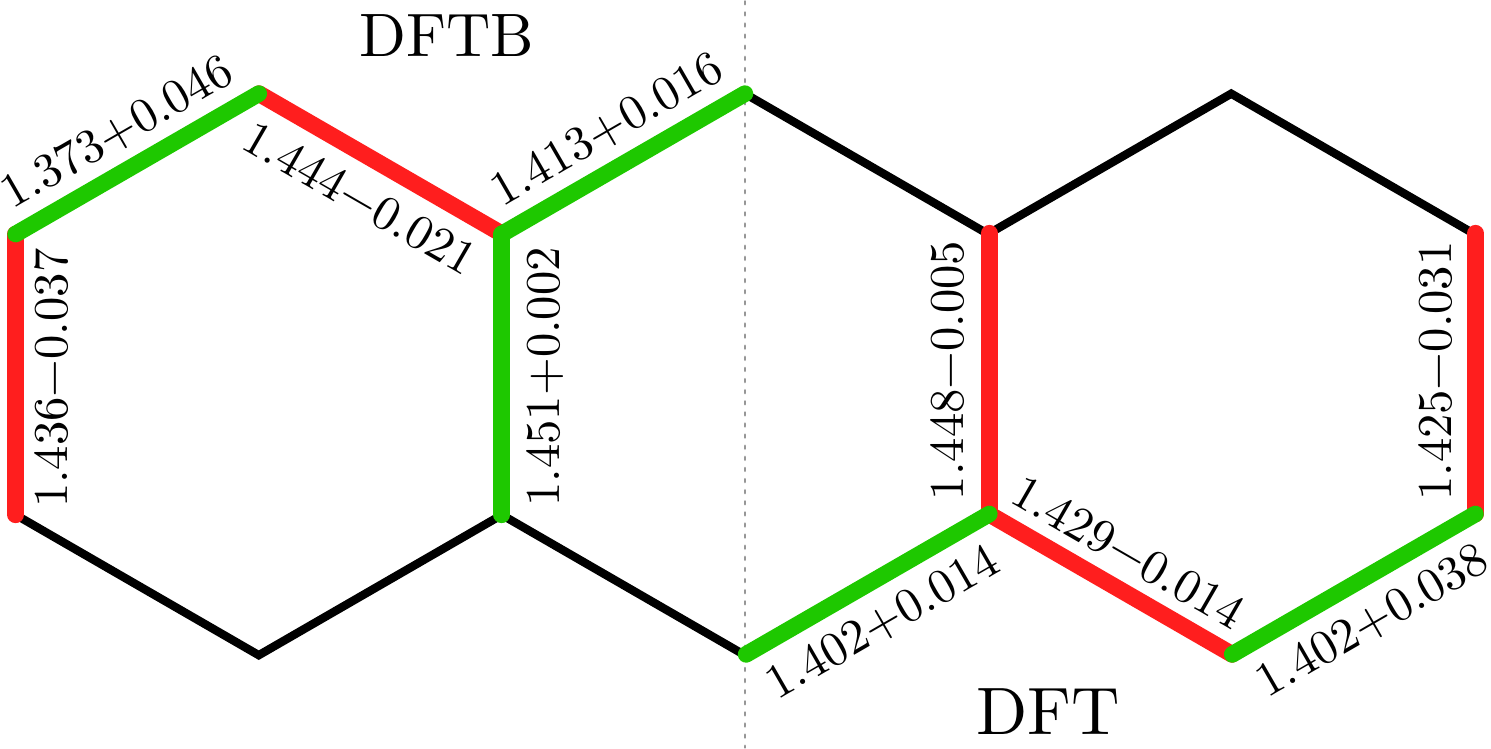}
\caption{\label{fig:anthracene}Deformation in the $\mathrm{S}_1$ state of anthracene calculated with DFTB (left) and DFT (right). Distances are given in \AA{}ngstr\"om.}
\end{figure}

As expected from experiment~\cite{ExpAnthracenePyrene} we observe a dipole-allowed $\mathrm S_0 \rightarrow \mathrm S_1$ transition ($L_a$ state in Platt nomenclature\cite{PlattNomenclature1949}) with $B_{2u}$ symmetry and an oscillator strength~$f = 0.048$ calculated with DFTB, and~$0.037$ with DFT.
The calculated vertical excitation energies are $\Delta (\vec R_\text{GS}) = 2.98\mathrm{eV}$ with DFTB, and $2.91\mathrm{eV}$ with DFT.
Optimization of the excited state leads to an equilibration of the bond lengths in the outer rings and a slight expansion of the outer bonds of the central ring for both DFTB and DFT, with a overall slightly stronger deformation for DFTB.
The DFT(B) ground state geometry and its deformation upon excitation is shown in figure~\ref{fig:anthracene}.
The deformation qualitatively agrees with the results obtained by~\citeauthor{DierksenVibronicDFT2004} using a hybrid functional~\cite{DierksenVibronicDFT2004}.
Comparison of the total energy of excited and ground state yields an $E_\text{0-0}$~energy of $2.65\mathrm{eV}$ for DFTB, and $2.67\mathrm{eV}$ for DFT, both of which considerably underestimate the experimentally obtained $E_\text{0-0}$ of $3.43\mathrm{eV}$ by almost $0.8\mathrm{eV}$~\cite{ExpAnthracenePyrene}.
It is known that density functional theory systematically underestimates energies of $L_a$~states in acenes~\cite{GrimmeParacAcenesTDDFT2003,ParacGrimmePAHWithTDDFT2003}, though this deficiency is corrected with range-separated hybrid functionals~\cite{PeachTDDFTEval2008}.

\begin{figure}[tbp]
\includegraphics[width=\columnwidth]{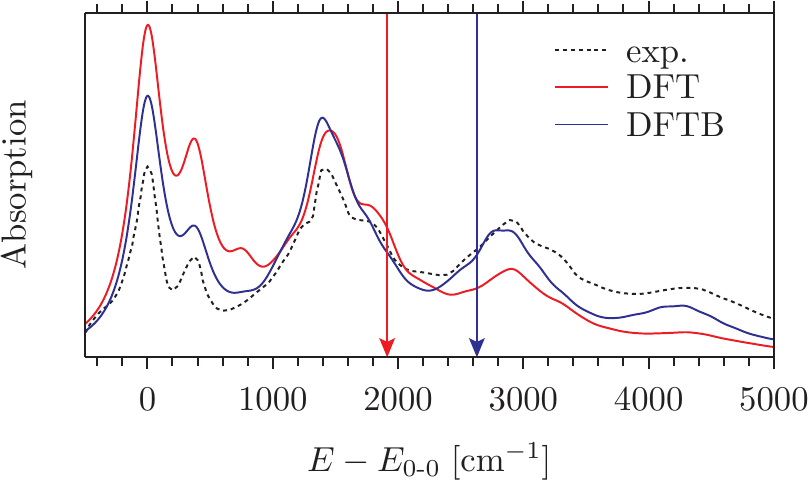}
\caption{\label{fig:anthracene_spectrum}Vibrationally resolved $\mathrm{S}_0 \rightarrow \mathrm{S}_1$ absorption spectrum of anthracene. The arrows mark the vertical excitation energies. Experimental spectrum from Ref.~\citenum{ExpAnthracenePyrene}.}
\end{figure}

The vibrational fine structure of the $\mathrm S_0 \rightarrow \mathrm S_1$ absorption band is shown in figure~\ref{fig:anthracene_spectrum}.
The general shape and the peak positions of both the DFT and DFTB calculated spectra agree very well with the experimental reference, though the experimental spectrum is overall a little more intense at higher energies.
This was also observed by \citeauthor{DierksenVibronicDFT2004} who hypothesized that this is due to a skewed baseline in the experimental spectrum~\cite{DierksenVibronicDFT2004}.
While they have later found that the effect reduces with a larger ratio of Hartree-Fock exchange (see also section~\ref{ss:pentarylene}), it never quite disappears as it does for other PAHs~\cite{DierksenVibronicEXX2004}.

Overall it can be said that the vibrational fine structure of the $\mathrm S_0 \rightarrow \mathrm S_1$ transition in anthracene is remarkably accurate compared the absolute energies, which are almost $0.8\mathrm{eV}$ too low.
This indicates that the excited state potential energy surface is mostly shifted to smaller energies, but not distorted in the process.
The similarity between the DFT and DFTB calculated spectra suggests that the additional approximations of DFTB have little influence on the quality of the obtained spectrum.

\subsection{Pentacene}

\begin{figure}[tbp]
\includegraphics[width=\columnwidth]{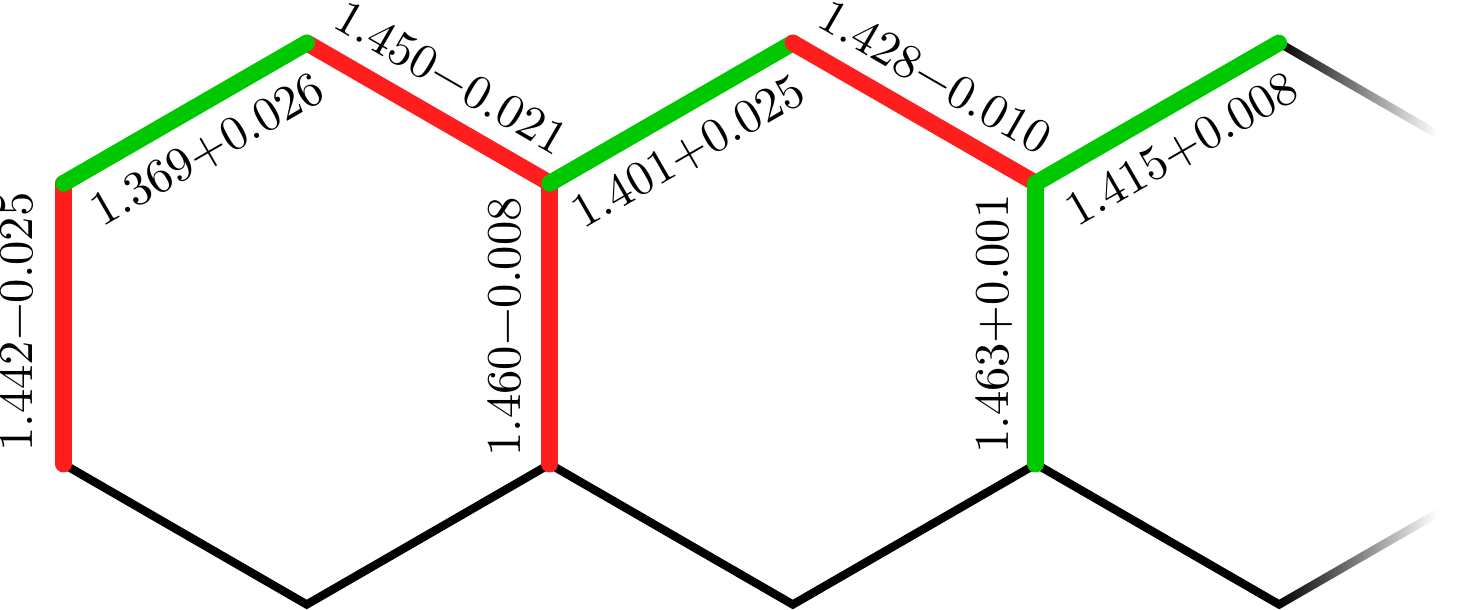}
\caption{\label{fig:pentacene}Deformation in the $\mathrm{S}_1$ state of pentacene as calculated with DFTB. Due to the symmetry of the deformation, only half of the molecule is shown. Distances are given in \AA{}ngstr\"om.}
\end{figure}

The next example is the $\mathrm S_0 \rightarrow \mathrm S_1$ transition in pentacene.
The $\mathrm S_1$ state is of $B_{2u}$ symmetry and labeled as $L_a$ in Platt's nomenclature~\cite{PlattNomenclature1949}.
The transition is dipole-allowed with an oscillator strength $f = 0.033$ calculated from DFTB and $0.059$ from DFT.
We have calculated a vertical excitation energy of $\Delta (\vec R_\text{GS}) = 1.80\mathrm{eV}$ with DFTB and $1.60\mathrm{eV}$ with DFT.
Optimization of the excited state yields an $E_\text{0-0}$~energy of $1.61\mathrm{eV}$ for DFTB and $1.48\mathrm{eV}$ for DFT, which is again too low when compared to the experimental $E_\text{0-0}$~energy of $2.12\mathrm{eV}$~\cite{ExpPentacene}.
The ground state bond distances and the deformation in the $\mathrm S_1$ state is shown in figure~\ref{fig:pentacene}.
As was the case for anthracene, the internal bonds are hardly affected in the excited state while bond distances on the outside tend to become more uniform, with larger deformations in the outer rings of the molecule.

\begin{figure}[tbp]
\includegraphics[width=\columnwidth]{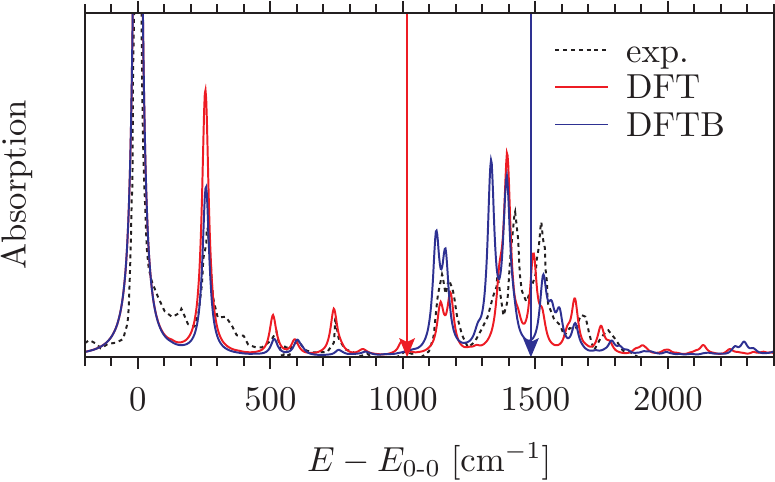}
\caption{\label{fig:pentacene_spectrum}Vibrationally resolved $\mathrm{S}_0 \rightarrow \mathrm{S}_1$ absorption spectrum of pentacene. The arrows mark the vertical excitation energies. Experimental spectrum from Ref.~\citenum{ExpPentacene}.}
\end{figure}

The vibrationally resolved $\mathrm S_0 \rightarrow \mathrm S_1$ absorption spectrum is shown in figure~\ref{fig:pentacene_spectrum}.
The experimental spectrum was measured in a $n$-hexadecane Shpol'skii matrix at a temperature of 5K and therefore has a much better spectral resolution than the vapor absorption spectrum we used for comparison in case of anthracene.
Both theoretical and experimental spectra show a strong absorption at the $E_\text{0-0}$~energy, indicating that there is a large probability for the nuclei to remain in their vibrational ground state during photon absorption.
Peak positions of the experimental spectrum are generally reproduced within $50\mathrm{cm}^{-1}$ by both DFTB and DFT.
Intensities are also generally well reproduced, with the exception of a line at $750\mathrm{cm}^{-1}$ which is almost missing with DFTB and a line at $1320\mathrm{cm}^{-1}$ that is too intense.
However, considering that embedding into different matrices yield slightly different experimental spectra~\cite{ExpPentacene}, while our calculation corresponds to absorption in the gas phase, both DFT and DFTB are in excellent agreement with the experimental data.

It is interesting to note that the vertical excitation energy (arrows in figure~\ref{fig:pentacene_spectrum}) is a rather poor approximation to the absorption maxima for both DFT and DFTB:
For DFTB it is clearly at too high energies and completely neglects the rather intense 0-0~transition.
For DFT it is closer to the ``mean absorption energy'' but paradoxically ends up at an energy where pentacene does not absorb at all.

\subsection{Pyrene}

\begin{figure}[tbp]
\includegraphics[width=0.8\columnwidth]{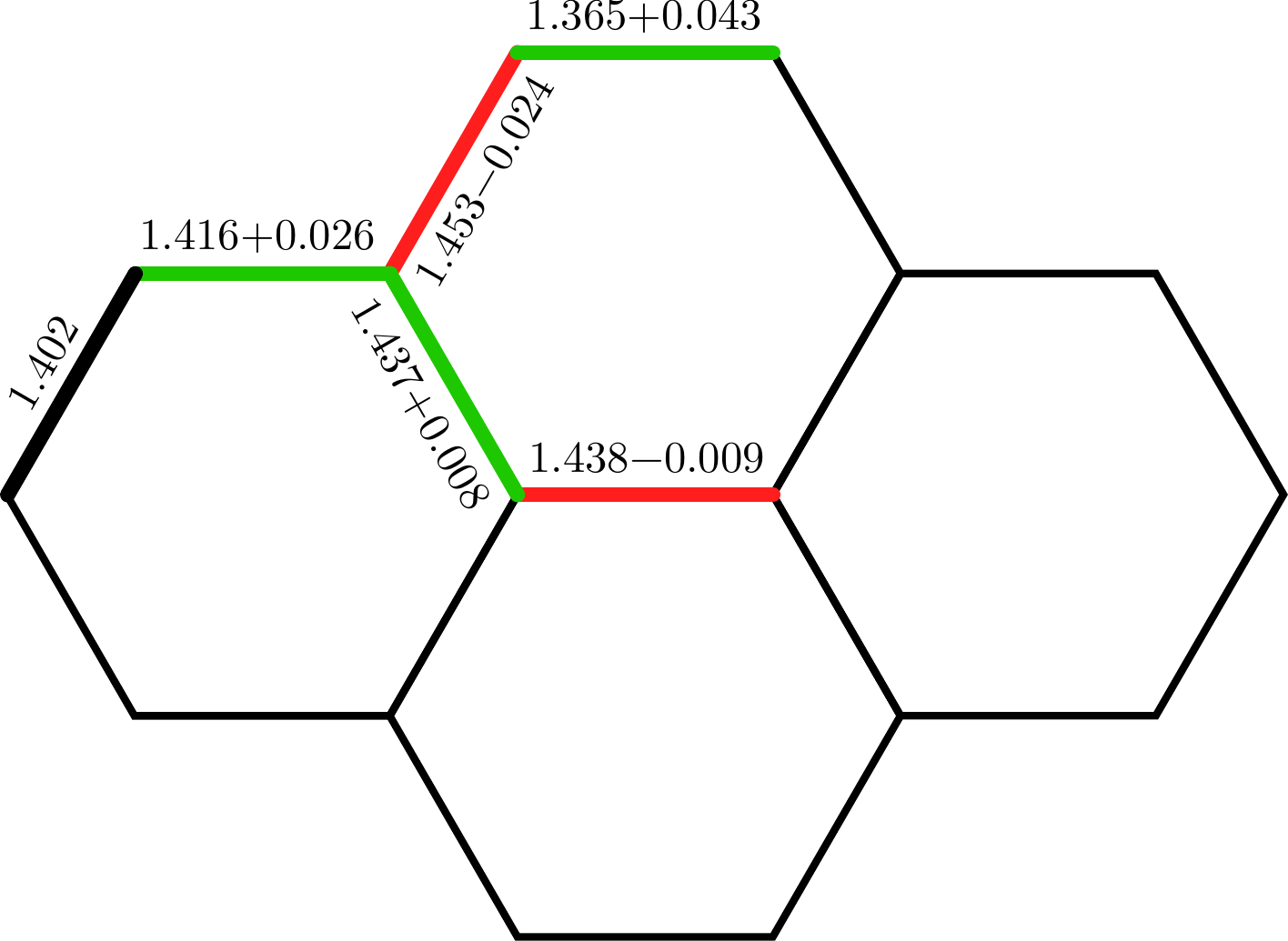}
\caption{\label{fig:pyrene}Deformation in the $\mathrm{S}_1$ state of pyrene as calculated with DFTB.  Distances are given in \AA{}ngstr\"om.}
\end{figure}

Another interesting test system is pyrene, which is not only a precursor to dyes such as pyranine, but as the smallest peri-fused PAH also structurally rather different from the previous acene examples.
Experimentally the absorption spectrum of pyrene shows a weak band associated with excitation into the~$L_b$ state with an $E_\text{0-0}$~energy of $3.36\mathrm{eV}$~\cite{GeiglePyreneLb1997} and an intense band of absorption into the~$L_a$ state with $E_\text{0-0} = 3.84\mathrm{eV}$~\cite{ExpAnthracenePyrene}.
Both DFT and DFTB erroneously predict the $L_a$~state to be the $\mathrm S_1$ state with vertical excitation energies of $\Delta (\vec R_\text{GS}) = 3.25\mathrm{eV}$ with DFTB and $3.38\mathrm{eV}$ with DFT.
The excitation into the $L_a$~state ($B_{3u}$ symmetry) is dipole allowed with an oscillator strengths of $f = 0.201$ for DFTB and $0.211$ for DFT.
The deformation of the molecule upon excitation is shown in figure~\ref{fig:pyrene} and is dominated by an equilibration of bond lengths along the perimeter of the molecule.
Comparing the total energies of ground and excited state we have found $E_\text{0-0}$~energies of $2.97\mathrm{eV}$ for DFTB and $3.17\mathrm{eV}$ with DFT, both of which are too low compared to experiment.

\begin{figure}[tbp]
\includegraphics[width=\columnwidth]{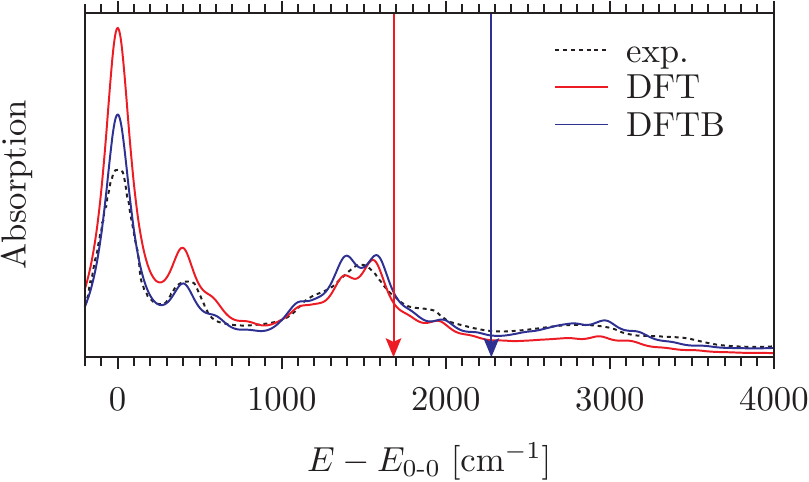}
\caption{\label{fig:pyrene_spectrum}Vibrational fine structure of the absorption into the $L_a$ state of pyrene. The arrows mark the vertical excitation energies. Experimental spectrum from Ref.~\citenum{ExpAnthracenePyrene}.}
\end{figure}

The vibrational fine structure of the absorption into the $L_a$ state ($\mathrm{S}_1$ theoretically, $\mathrm{S}_2$ experimentally) is shown in figure~\ref{fig:pyrene_spectrum}.
The agreement between the theoretical spectra and experiment is almost perfect.
This indicates that with DFT and DFTB the excited state potential energy surfaces are not distorted but merely shifted in energy, which is interesting considering that the shift is quite severe and even reverses the energetic ordering of $L_a$ and $L_b$~state with respect to experiment.

\subsection{\label{ss:pentarylene}Pentarylene}

The examples so far were rather small molecules where even for DFT the calculation of the vibrational structure of the absorption band belonging to a single electronic excitation is computationally not a problem.
For larger molecules or if multiple electronic states have to be considered, performance will be come an issue though.
In order to investigate these computational aspects, we have calculated the vibrational fine structure of the $\mathrm S_0 \rightarrow \mathrm S_1$ transition in pentarylene.
With 74 atoms, pentarylene is much larger than the previous example molecules. 
 
\begin{figure}[tbp]
\includegraphics[width=\columnwidth]{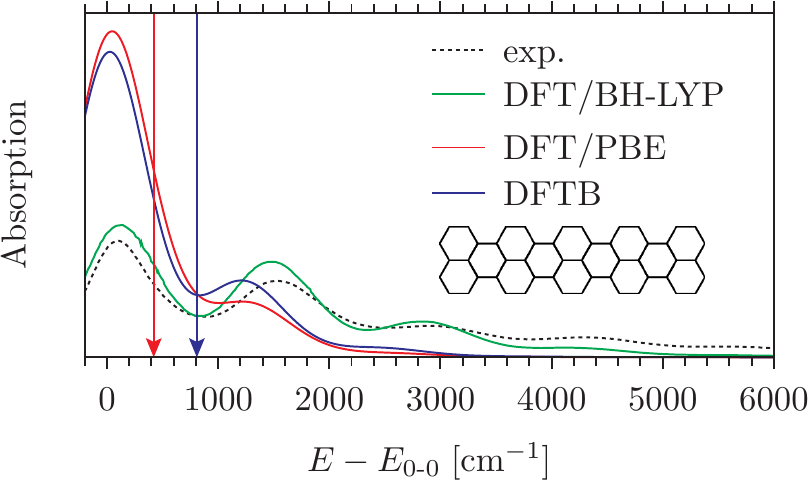}
\caption{\label{fig:pentarylene_spectrum}Vibrationally resolved $\mathrm{S}_0 \rightarrow \mathrm{S}_1$ absorption spectrum of pentarylene. The arrows mark the vertical excitation energies. DFT/BH-LYP results from Ref.~\citenum{DierksenVibronicEXX2004}. Experimental spectrum from Ref.~\citenum{ExpPentarylene}.}
\end{figure}

The $\mathrm S_1$ state in pentarylene has $B_{3u}$ symmetry and the transition from the ground state is strongly dipole allowed with oscillator strengths of $f = 1.58$ with DFTB and $1.48$ with DFT.
The calculated $E_\text{0-0}$ energies ($1.38\mathrm{eV}$ for DFTB, $1.31\mathrm{eV}$ for DFT) again underestimate the experimental value of $1.66\mathrm{eV}$~\cite{ExpPentarylene}. The vibrational structure of the absorption band is shown in figure~\ref{fig:pentarylene_spectrum}.
The agreement of the calculated spectra with experiment is rather bad: Both DFT/PBE and DFTB show a dominating 0-0 transitions, while the experimental spectrum is fairly wide and features at least 4 distinct maxima of decreasing intensity~\cite{ExpPentarylene}. This difference indicates that the geometry deformation between ground and excited state is underestimated with both DFT/PBE and DFTB, making the overlap between the nuclear ground state wavefunctions too large and the 0-0 transition too likely. This was also observed by \citeauthor{DierksenVibronicEXX2004}, who have found that a large ratio of exact exchange (50\% with the BH-LYP functional) is needed in order to reproduce the experimental spectrum~\cite{DierksenVibronicEXX2004}. However, even though DFT/PBE and DFTB both disagree with experiment, they agree very well with each other, indicating that the deficiency is already present at the level of DFT with GGA functionals and was not introduced with the additional approximations in DFTB.

Looking at the computational performance, it is clear that the calculation of vibrational frequencies in the excited state is the bottleneck for these calculations. With the numerical differentiation of the analytical gradient, $6N_\text{atom}$ single point TD-DFT(B) calculations are necessary to determine the Hessian. Running on an Intel Core i7-4770 CPU the entire calculation took 49~hours for DFT, out of which 43~hours were spent on the excited state Hessian. With DFTB it is also the evaluation of the Hessian that takes the most time, but the entire calculation finishes within 7~minutes, which is a speedup by a factor of 420 when compared to DFT.

\subsection{Octatetraene}

In addition to the aromatic compounds covered in the previous sections, polyenes are another class of systems known for their optical properties: On the one hand they exhibit intensely absorbing $\pi \rightarrow \pi^*$ transitions, the energy of which can be tuned through the number of conjugated double bonds, making them excellent dyes. On the other hand they can also undergo \textit{cis}-\textit{trans}~isomerization upon absorption of a photon; a process critical in the biochemical conversion of radiative into chemical energy.

\begin{figure}[tbp]
\includegraphics[width=\columnwidth]{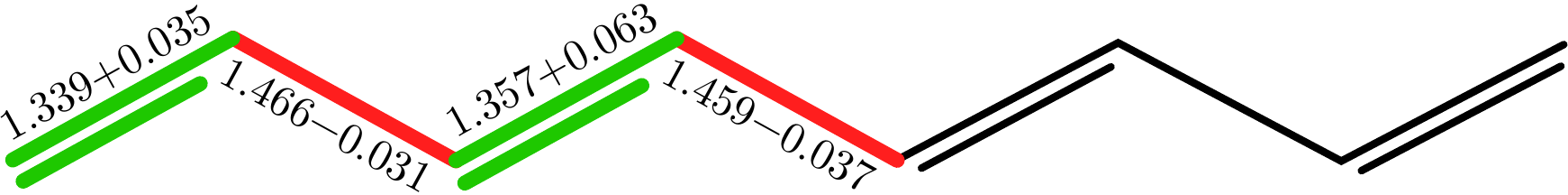}
\caption{\label{fig:octatetraene}Deformation in the $1^1B_u$ state of octatetraene as calculated with DFTB.  Distances are given in \AA{}ngstr\"om.}
\end{figure}

Our specific example is the excitation into $1^1B_u$ state in all-$trans$ octatetraene. This transition is strongly dipole allowed with an oscillator strength of $f = 1.14$ with DFTB, and $1.40$ with DFT. For both DFTB and DFT this is the $\mathrm S_1$~state with vertical excitation energies of $\Delta (\vec R_\text{GS}) = 3.83\mathrm{eV}$ for DFTB and $3.78\mathrm{eV}$ for DFT. However, experimentally a weak dipole-forbidden absorption into the $2^1A_g$ state is observed at $E_\text{0-0} = 3.60\mathrm{eV}$~\cite{Exp2Octatetraene}, while absorption into the $1^1B_u$~state only starts at $E_\text{0-0} = 4.40\mathrm{eV}$~\cite{ExpOctatetraene}. Optimization of the $1^1B_u$~excited state leads to an equilibration of bond lengths, in which double bonds stretch and single bonds contract. This is shown in figure~\ref{fig:octatetraene}.
The calculated $E_\text{0-0}$~energies are $3.44\mathrm{eV}$ with DFTB and $3.53\mathrm{eV}$ with DFT, both of which are almost $1\mathrm{eV}$ too low compared to experiment.

\begin{figure}[tbp]
\includegraphics[width=\columnwidth]{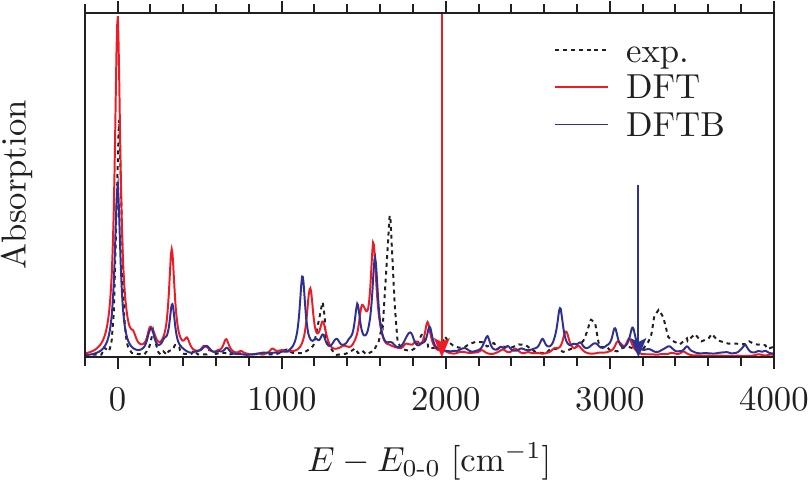}
\caption{\label{fig:octatetraene_spectrum}Vibrational fine structure of the absorption into the $1^1B_u$ state of octatetraene. The arrows mark the vertical excitation energies. Experimental spectrum from Ref.~\citenum{ExpOctatetraene}}
\end{figure}

The vibrationally resolved absorption into the $1^1B_u$~state is shown in figure~\ref{fig:octatetraene_spectrum}.
While both DFTB and DFT predict the experimentally seen, very intense 0-0 transition, the rest of the spectrum is rather poorly described. The most pronounced peaks in the experimental spectrum are found at $1235\mathrm{cm}^{-1}$ and $1645\mathrm{cm}^{-1}$. Additional absorption peaks are seen at $2880\mathrm{cm}^{-1} = (1235 + 1645)\mathrm{cm}^{-1}$ and $3290\mathrm{cm}^{-1} = 2 \times 1645\mathrm{cm}^{-1}$, corresponding to combined and double excited modes respectively~\cite{ExpOctatetraene}. Compared to the experimental spectrum, both calculated spectra show the prominent peaks at slightly lower energies: With DFTB the corresponding modes are found at $1125\mathrm{cm}^{-1}$ and $1570\mathrm{cm}^{-1}$, with the combined and double nuclear excitations visible at $2695\mathrm{cm}^{-1}$ and $3140\mathrm{cm}^{-1}$. Additionally there are rather intense peaks at $331\mathrm{cm}^{-1}$ and $1460\mathrm{cm}^{-1}$, which were not observed in the experimental spectrum. However, the spectra obtained with DFTB and DFT are almost identical, indicating that these deficiencies are already present at the DFT/GGA level and have not been introduced through the additional approximations in DFTB. As was the case for the PAHs, \citeauthor{DierksenVibronicEXX2004} have found that the DFT results can be improved by inclusion of exact exchange in the exchange-correlation functional, though for octatetraene they have found that an especially large amount of exact exchange (>50\%) is required for a better agreement with the experimental data.

\subsection{Stilbene}

The optical properties of stilbene are used in various technical applications such as dye lasers, optical brighteners and as a scintillator material.
In addition to its technical importance it is also an interesting test system due to the fact that it contains both a conjugated double bond and aromatic rings.

\begin{figure}[tbp]
\includegraphics[width=0.9\columnwidth]{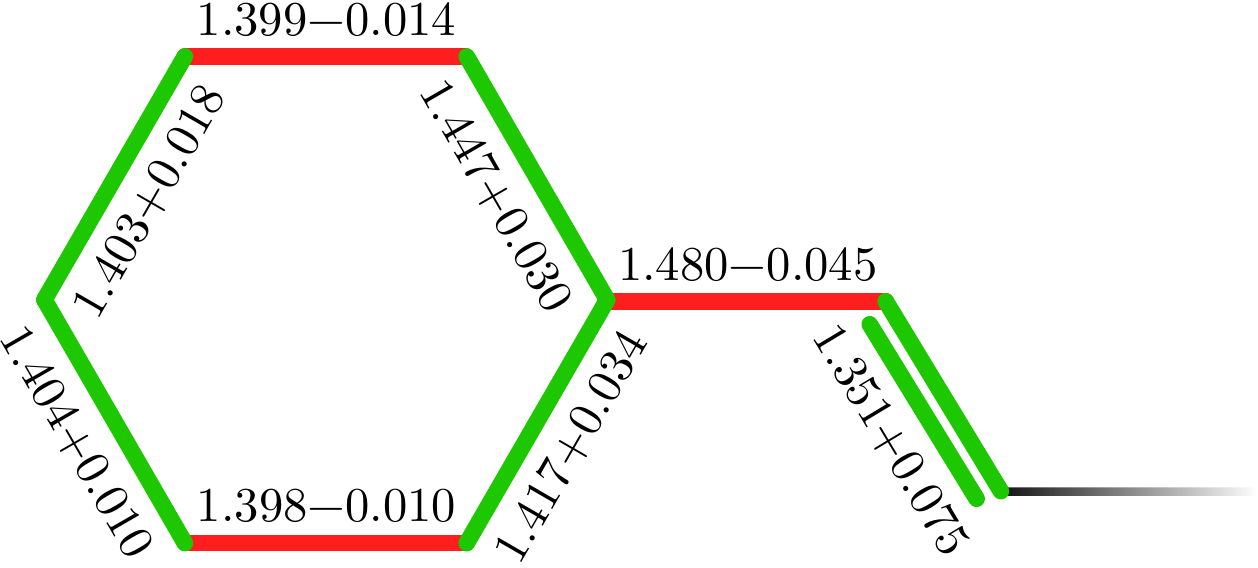}
\caption{\label{fig:stilbene}Deformation in the $\mathrm{S}_1$ state of \textit{trans}-stilbene as calculated with DFTB. Distances are given in \AA{}ngstr\"om.}
\end{figure}

We have studied the $\mathrm S_0 \rightarrow \mathrm S_1$ excitation in \textit{trans}-stilbene.
The $\mathrm S_1$ state has $B_{u}$ symmetry and the absorption is a strongly dipole allowed $\pi \rightarrow \pi^*$ transition with an oscillator strengths of $f = 0.83$ for DFTB and $0.85$ for DFT.
The calculated vertical excitation energies of $\Delta (\vec R_\text{GS}) = 3.74\mathrm{eV}$ from DFTB and $3.60\mathrm{eV}$ agree surprisingly well with the experimental $E_\text{0-0}$~energy of $3.80\mathrm{eV}$~\cite{ExpStilbene}, though the experimental value was measured in a methyl pentane solution, so a solvent induced shift should be kept in mind. Optimization of the excited state shows the strongest change in geometry at the central double bond and smaller displacements further away, see figure~\ref{fig:stilbene}.
The calculated $E_\text{0-0}$ energies of $3.34\mathrm{eV}$ with DFTB and $3.32\mathrm{eV}$ with DFT are too small compared to the experimental value of $3.80\mathrm{eV}$, though the relative error is small compared to some other examples.

\begin{figure}[tbp]
\includegraphics[width=\columnwidth]{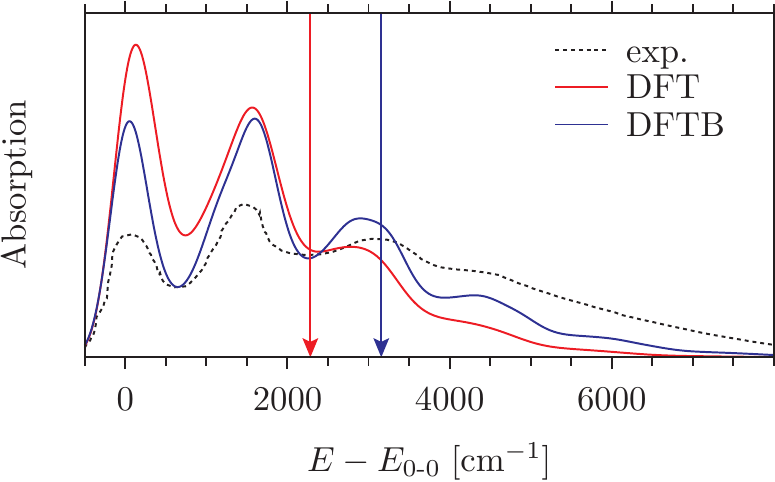}
\caption{\label{fig:stilbene_spectrum}Vibrationally resolved $\mathrm{S}_0 \rightarrow \mathrm{S}_1$ absorption spectrum of \textit{trans}-stilbene. The arrows mark the vertical excitation energies. Experimental spectrum from Ref.~\citenum{ExpStilbene}.}
\end{figure}

The vibrational fine structure of the $\mathrm S_0 \rightarrow \mathrm S_1$ transition in \textit{trans}-stilbene is shown in figure~\ref{fig:stilbene_spectrum}.
While peak positions are relatively well described, both DFT and DFTB overestimate the intensity of the 0-0 transition, indicating that the geometric displacement between ground and excited state is underestimated.
The same problem was also observed for the pentarylene example, where it was found that a larger amount of exact exchange improves the results~\cite{DierksenVibronicEXX2004}. This is also the case for the $\mathrm S_0 \rightarrow \mathrm S_1$ transition in \textit{trans}-stilbene, where \citeauthor{DierksenVibronicEXX2004} have found that 30--40$\%$ of exact exchange give the best agreement with experiment.

\subsection{Anisole}

Moving away from pure hydrocarbons, we have calculated the excitation into the $1^1A'$~state of anisole.
The $1^1A'$~state is the $S_1$ at the ground state's equilibrium geometry and its excitation is dipole allowed with an oscillator strength of $0.03$ and $0.036$ for DFT and DFTB, respectively.
With $E_\text{0-0}$ energies of $4.22\mathrm{eV}$ with DFTB and $4.38\mathrm{eV}$ with DFT, both methods slightly underestimate the experimental 0-0~energy of $4.51\mathrm{eV}$.
Compared to the ground state geometry, all bonds in the benzene ring expand upon excitation. It is interesting to note that this expansion is larger for DFTB, with a maximum bond elongation of~$7\mathrm{pm}$ compared to~$4\mathrm{pm}$ with DFT.
Furthermore the C-N bond within the methoxy group elongates slightly upon excitation in DFT, while it shrinks by $2\mathrm{pm}$ for DFTB.

\begin{figure}[tbp]
\includegraphics[width=\columnwidth]{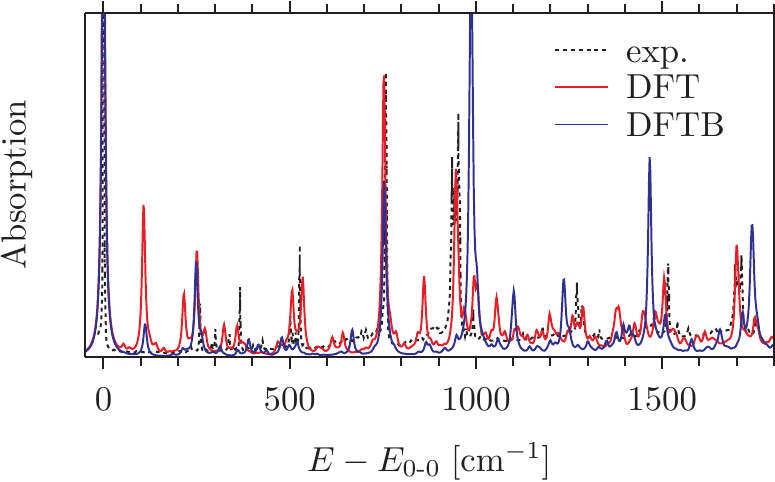}
\caption{\label{fig:anisole_spectrum}Vibrationally resolved $\mathrm{S}_0 \rightarrow \mathrm{S}_1$ absorption spectrum of anisole. Vertical excitation energies are outside of the plotted spectral range. The experimental spectrum from Ref.~\citenum{AnisoleExp} has been scaled to match the intensity of the feature at~$750\mathrm{cm}^{-1}$ with the one calculated from DFT.}
\end{figure}

The vibrational fine structure of the transition is shown in figure~\ref{fig:anisole_spectrum}.
With the exception of two peaks at $100\mathrm{cm}^{-1}$ and $860\mathrm{cm}^{-1}$, DFT at the GGA level reproduces the experimental spectrum~\cite{AnisoleExp} very well.
(A spectrum calculated with the B3LYP hybrid functional can be found in Ref.~\citenum{BaroneFrankCondon2009} and is almost identical to the GGA calculation.)
DFTB qualitatively reproduces the experimental spectrum but compared to DFT shows larger deviations in peak positions.
Peak intensities also seem to be worse with DFTB at first sight.
However, one should keep in mind that the experimental data has been scaled to match the intensity of the feature at~$750\mathrm{cm}^{-1}$ to the DFT spectrum.
Nevertheless, considering the entire spectral range (including the region~$>1800\mathrm{cm}^{-1}$ not shown in figure~\ref{fig:anisole_spectrum}), DFTB predicts larger intensities further away from the 0-0~origin.
This is consistent with the larger geometric deformation seen in the excited state with DFTB, which due to the displaced minima requires more vibrational quanta in the excited state to reach overlap with the original nuclear wavefunction.

\subsection{Coumarin dye C480}

\begin{figure}[tbp]
\includegraphics[width=\columnwidth]{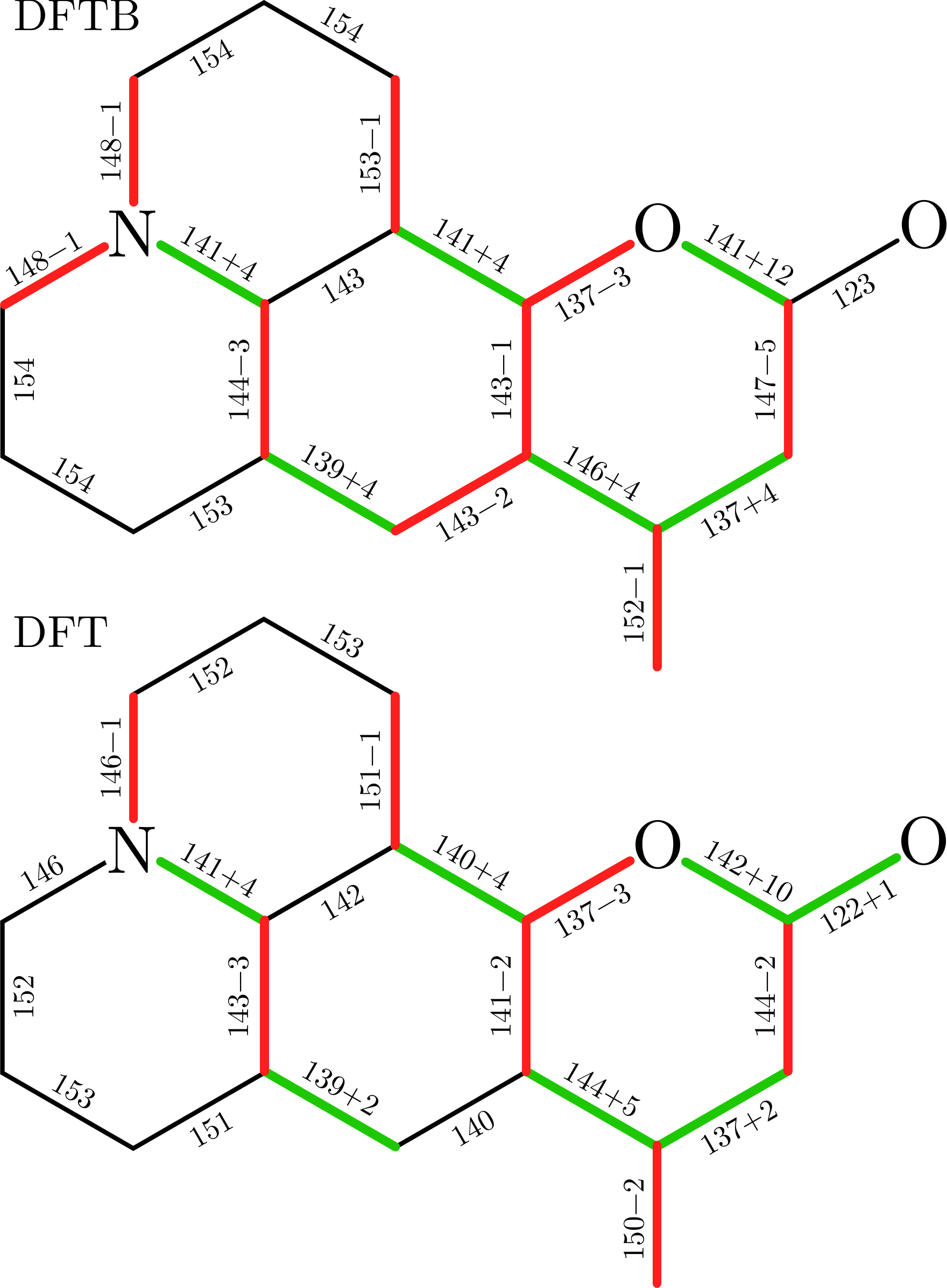}
\caption{\label{fig:C480_deformation}Deformation in the $\mathrm{S}_1$ state of the coumarin dye~C480 as calculated with DFTB (top) and DFT (bottom). Distances are given in picometers.}
\end{figure}

\begin{figure}[tbp]
\includegraphics[width=\columnwidth]{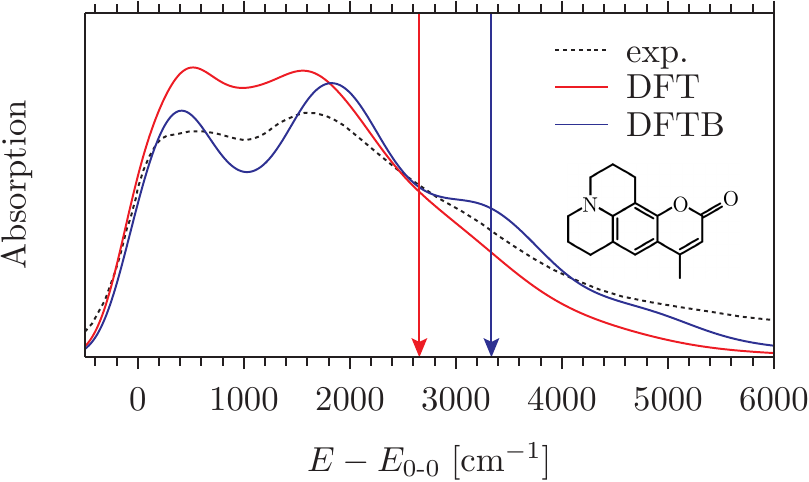}
\caption{\label{fig:C480_spectrum}Vibrationally resolved $\mathrm{S}_0 \rightarrow \mathrm{S}_1$ absorption spectrum of the coumarin dye~C480. The arrows mark the vertical excitation energies. Experimental spectrum from Ref.~\citenum{BaroneVibResSpecComparison2015}.}
\end{figure}

As an example of a heterocyclic compound we have chosen the coumarin dye C480 (structure inlayed in figure~\ref{fig:C480_spectrum}).
As a typical dye molecule, coumarin C480 has a strongly dipole allowed $\mathrm{S}_0 \rightarrow \mathrm{S}_1$~transition with $\pi \rightarrow \pi^*$~character (HOMO $\rightarrow$ LUMO).
We have calculated 0-0~energies of $2.73\mathrm{eV}$ and $2.77\mathrm{eV}$ with DFT and DFTB, respectively.
As for all other compounds, both methods underestimate the experimental 0-0~energy of $\approx 3.22\mathrm{eV}$ determined from the inset of the first band in the absorption spectrum measured in methylcyclohexane~\cite{BaroneVibResSpecComparison2015}.
The deformation upon excitation is shown in figure~\ref{fig:C480_deformation} and is mostly restricted to the core coumarin and the C-N~bond linking the nitrogen atom to the coumarin unit.
Especially noteworthy is the very strong elongation ($+12\mathrm{pm}$ for DFTB; $+10\mathrm{pm}$ for DFT) of the C-O~bond between the heterocyclic oxygen and the carbonyl carbon.
The deformation upon excitation is very similar for DFT and DFTB, although, as was already observed in previous examples, DFTB predicts overall slightly larger deformations.

The vibrational fine structure of the $\mathrm S_0 \rightarrow \mathrm S_1$ transition in C480 is shown in figure~\ref{fig:C480_spectrum}.
The DFT calculated spectrum agrees very well with the experimental absorption spectrum measured in methylcyclohexane~\cite{BaroneVibResSpecComparison2015}, though contrary to experiment, DFT predicts the first absorption maximum to be slightly more intense than the second.
For DFTB the relative intensities of the two maxima agree with experiment, but the absorption band is overall slightly wider, which we attribute to the larger geometric deformations seen in DFTB.

The coumarin C480 dye has recently also been used by~\citeauthor{BaroneVibResSpecComparison2015} in a benchmark study~\cite{BaroneVibResSpecComparison2015} investigating the performance of different (global and range-separated hybrid) exchange-correlation functionals on the vibronic structure and absolute band positions.
The authors found that the $\omega$B97X~functional~\cite{ChaiOmegaB97X2008} predicts the best spectral shapes but that no single functional simultaneously provides accurate band positions and shapes.
In fact $\omega$B97X overestimates the $E_\text{0-0}$~energy by about as much as we have found PBE to underestimate it.
GGA functionals were unfortunately not included in the comparison in Ref.~\citenum{BaroneVibResSpecComparison2015}, but considering our results, the system seems to be sufficiently well described at the GGA level and by extension with DFTB.

\subsection{Bithiophene}

As another example for a heterocyclic compound we have calculated the $\mathrm{S}_1 \rightarrow \mathrm{S}_0$ fluorescence spectrum of planar \textit{trans}-2,2'-bithiophene.
The fluorescence of bithiophene has recently also been used in a benchmark study~\cite{StendardoVibResDithiophene2012} by \citeauthor{StendardoVibResDithiophene2012} investigating the effect of different exchange-correlation functionals on the vibrational fine structure of the emission line.
The \texttt{3ob:freq}~parameter set used so far does not include parameters for sulfur. However, different versions of the \texttt{3ob}~set are cross compatible among each other, so that sulfur parameters from a newer version (with sulfur repulsive potentials not specifically optimized for frequency calculations) were used.
For planar bithiophene the $\mathrm{S}_1$ has $1^1B_u$~symmetry and we have calculated 0-0~energies of $3.38\mathrm{eV}$ and $3.30\mathrm{eV}$ with DFT and DFTB, respectively.
As with all examples so far, this underestimates the experimental $E_\text{0-0}$ of $3.86\mathrm{eV}$.

\begin{figure}[tbp]
\includegraphics[width=\columnwidth]{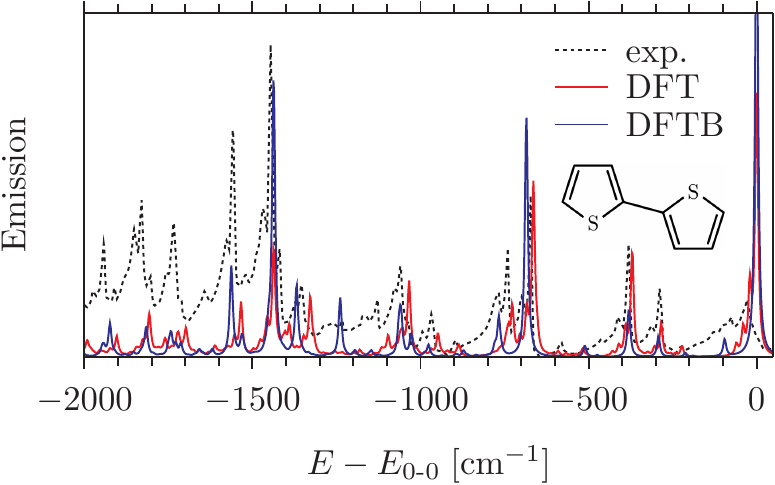}
\caption{\label{fig:dithiophene_spectrum}$\mathrm{S}_1 \rightarrow \mathrm{S}_0$ fluorescence spectrum of bithiophene. The experimental spectrum from Ref.~\citenum{ExpDithiophene} has been scaled to match the intensity of the feature at~$390\mathrm{cm}^{-1}$ with the one calculated from DFT.}
\end{figure}

The fluorescence spectrum of bithiophene is shown in figure~\ref{fig:dithiophene_spectrum}.
Both DFT and DFTB show very good agreement in peak positions compared to experiment~\cite{ExpDithiophene}.
The biggest difference between the DFT and DFTB calculated spectra is the relative intensity of the two intense features at $700\mathrm{cm}^{-1}$ and $1450\mathrm{cm}^{-1}$.
Here DFT predicts a higher intensity of the $700\mathrm{cm}^{-1}$~peak, while DFTB predicts the opposite.
\citeauthor{StendardoVibResDithiophene2012} have pointed out that the relative intensity of the two features is extremely sensitive to the choice of the functional, and furthermore depends strongly on the experimental conditions, with the $1450\mathrm{cm}^{-1}$~peak being much more intense in a hexane matrix~\cite{ExpDithiophene} than in a jet-cooled beam~\cite{ExpDithiophene2}.
It appears that both DFT and DFTB underestimate to overall intensity in the spectral region~$<-1500\mathrm{cm}^{-1}$.
However, the experimental spectrum has been scaled to match the intensity of the feature at~$390\mathrm{cm}^{-1}$ with the one calculated from DFT, so absolute intensities should not be overinterpreted.

\subsection{Triazoline}

\begin{figure}[b]
\includegraphics[width=0.9\columnwidth]{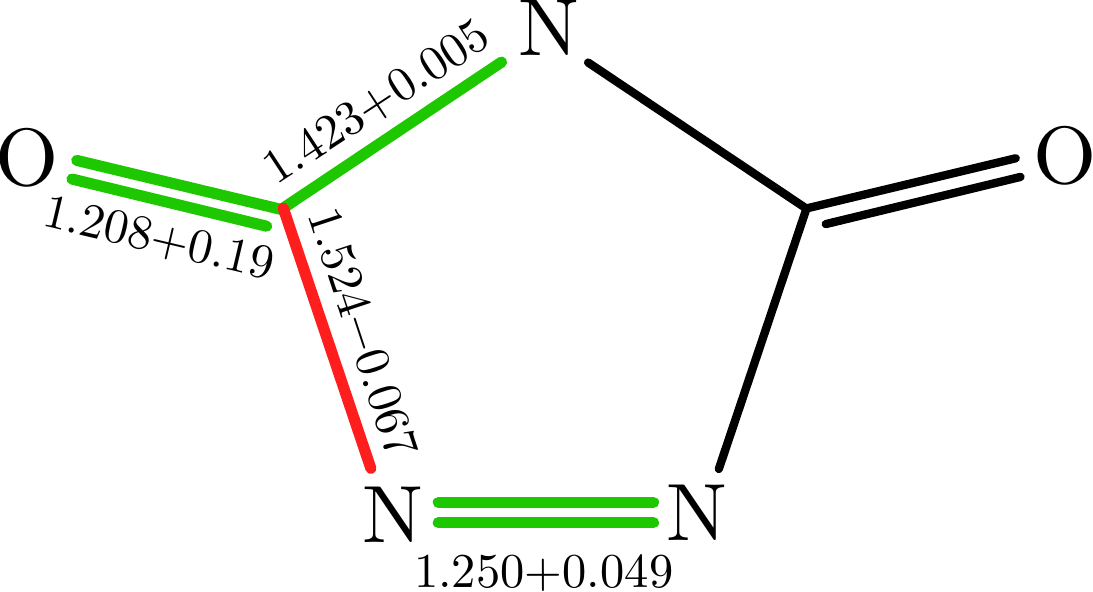}
\caption{\label{fig:triazoline}Deformation in the $\mathrm{S}_1$ state of triazoline as calculated with DFTB. Distances are given in \AA{}ngstr\"om.}
\end{figure}

The last example we want to look at is the $\mathrm S_0 \rightarrow \mathrm S_1$ excitation in 4-H-1,2,4-triazoline-3,5-diones.
As opposed to the other examples, which were $\pi \rightarrow \pi^*$ transitions, the $\mathrm S_0 \rightarrow \mathrm S_1$ transition in triazoline has $n \rightarrow \pi^*$ character.
This is an interesting test case as TD-DFTB is known to fail for $\sigma \rightarrow \pi^*$ and $n \rightarrow \pi^*$ transitions in that it predicts zero oscillator strengths and vanishing singlet-triplet gaps~\cite{NiehausTDDFTBOnsiteAndFracOcc2013}.
We therefore expect to see significant differences in the spectra calculated with (TD-)DFTB and (TD-)DFT at the GGA level, which is in contrast to the close correspondence we have observed for the other systems.
Note that TD-DFTB's failure for these transitions has recently been corrected by \citeauthor{NiehausTDDFTBOnsiteAndFracOcc2013} through inclusion of one-center integrals of the exchange type~\cite{NiehausTDDFTBOnsiteAndFracOcc2013}.
However, this so-called on-site correction to TD-DFTB is fairly involved and analytical excited state gradients are not yet available.
We will hence restrict our discussion to TD-DFTB in its original formulation~\cite{NiehausTDDFTB2001}.

The $\mathrm S_0 \rightarrow \mathrm S_1$ excitation in triazoline is very weakly dipole-allowed with an oscillator strength of $f = 0.0006$ for DFT, while DFTB mispredicts the oscillator strength to be exactly zero due to the above mentioned problem.
Optimization of the excited state with DFTB leads to a stretching of the N-N double bond and a shrinking of the adjacent C-N bond. This is shown in figure~\ref{fig:triazoline}.
The calculated $E_\text{0-0}$~energies are $1.60\mathrm{eV}$ with DFTB and $1.42\mathrm{eV}$ with DFT, both of which are considerably too low compared to the experimental value of $2.15\mathrm{eV}$~\cite{ExpTriazoline}.

\begin{figure}[tbp]
\includegraphics[width=\columnwidth]{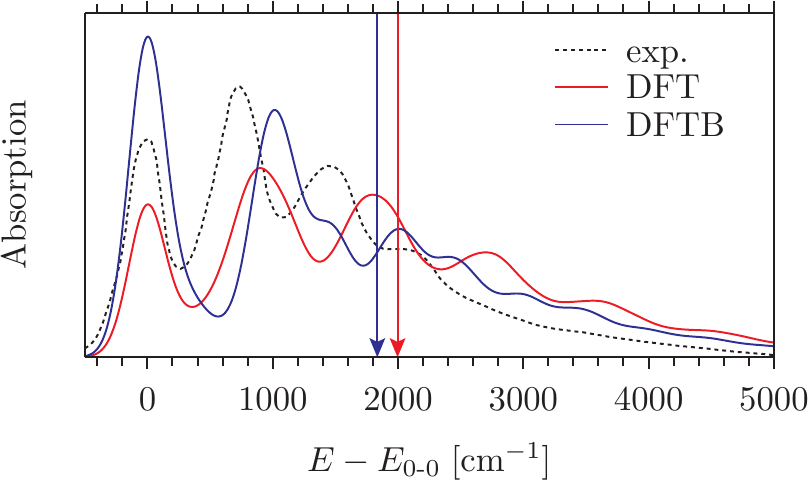}
\caption{\label{fig:triazoline_spectrum}Vibrationally resolved $\mathrm{S}_0 \rightarrow \mathrm{S}_1$ absorption spectrum of triazoline. The arrows mark the vertical excitation energies. Experimental spectrum from Ref.~\citenum{ExpTriazoline}.}
\end{figure}

The vibrational structure of the absorption band is shown in figure~\ref{fig:triazoline_spectrum}.
DFTB indeed predicts a qualitatively wrong spectrum in which the 0-0 transition has the highest intensity, while both DFT and the experimental spectrum have their absorption maximum around $1000\mathrm{cm}^{-1}$.
The spectrum obtained with DFT is overall too wide when compared to experiment, but according to \citeauthor{DierksenVibronicEXX2004} this problem can be resolved by using hybrid exchange-correlation functionals~\cite{DierksenVibronicEXX2004}.

\section{\label{s:conclusion}Conclusion}

We have shown that DFTB is an excellent and computationally very efficient approximation to DFT at the GGA level for the calculation of the vibronic fine structure of UV/Vis absorption bands.
Using the recent \texttt{3ob:freq} parameter set~\cite{Elstner3obParameters2013,Elstner3obSPParameters2014} very good agreement with DFT calculated spectra was achieved at a fraction of the computational cost.

We have found that experimentally measured vibrational fine structures are often reproduced by GGA DFT and DFTB, even when absolute excitation energies are in significant error compared to experiment.
This shows that the shape of the excited state potential energy surface is well reproduced in both GGA DFT and DFTB, even though the surface may be shifted in energy.

In cases such as pentarylene and stilbene, where the experimentally seen vibronic structure is not well reproduced by DFT at the GGA level and DFTB, better agreement with experiment can usually be obtained using hybrid exchange-correlation functionals~\cite{DierksenVibronicEXX2004}.
While this is relatively straightforward and well established in DFT, work on including exact exchange in the DFTB framework has only recently begun~\cite{NiehausLCDFTB2011,LutskerLCDFTB2015,HumeniukLCTDDFTB2015} and analytical TD-DFTB gradients are not yet available for these extensions.

The only example where we found a large discrepancy between DFT and DFTB was the $\mathrm S_0 \rightarrow \mathrm S_1$ excitation in triazoline. This is caused by the known failure of TD-DFTB for $n \rightarrow \pi^*$~transitions which was recently removed by~\citeauthor{NiehausTDDFTBOnsiteAndFracOcc2013} with the so-called on-site correction~\cite{NiehausTDDFTBOnsiteAndFracOcc2013}. However, since analytical gradients are not yet available for on-site corrected DFTB, it can presently not be used efficiently for the calculation of vibronic fine structures.

In summary, we believe that the good performance of (TD-)DFTB for the calculation of vibronic effects in UV/Vis spectra makes the inclusion of these effects possible for applications where they would previously have been neglected due to their computational cost.
Care should be taken when the method is applied to excitations with $\sigma \rightarrow \pi^*$ or $n \rightarrow \pi^*$ character or excitations that are not well described by (TD-)DFT at the GGA level.
However, both these restrictions are likely to be lifted with recent DFTB extensions, i.e.\ on-site correction~\cite{NiehausTDDFTBOnsiteAndFracOcc2013} and inclusion of exact exchange~\cite{NiehausLCDFTB2011,LutskerLCDFTB2015,HumeniukLCTDDFTB2015}, which due to the lack of analytical (excited state) gradients, can not yet be used for the calculation of vibronic fine structure.

\begin{acknowledgements}
The research leading to these results has received funding from the European Union's Seventh Framework Programme (FP7-PEOPLE-2012-ITN) under project PROPAGATE, GA~316897.
\end{acknowledgements}

\bibliography{literature,paper2,paper1}

\begin{thebibliography}{62}%
\makeatletter
\providecommand \@ifxundefined [1]{%
 \@ifx{#1\undefined}
}%
\providecommand \@ifnum [1]{%
 \ifnum #1\expandafter \@firstoftwo
 \else \expandafter \@secondoftwo
 \fi
}%
\providecommand \@ifx [1]{%
 \ifx #1\expandafter \@firstoftwo
 \else \expandafter \@secondoftwo
 \fi
}%
\providecommand \natexlab [1]{#1}%
\providecommand \enquote  [1]{``#1''}%
\providecommand \bibnamefont  [1]{#1}%
\providecommand \bibfnamefont [1]{#1}%
\providecommand \citenamefont [1]{#1}%
\providecommand \href@noop [0]{\@secondoftwo}%
\providecommand \href [0]{\begingroup \@sanitize@url \@href}%
\providecommand \@href[1]{\@@startlink{#1}\@@href}%
\providecommand \@@href[1]{\endgroup#1\@@endlink}%
\providecommand \@sanitize@url [0]{\catcode `\\12\catcode `\$12\catcode
  `\&12\catcode `\#12\catcode `\^12\catcode `\_12\catcode `\%12\relax}%
\providecommand \@@startlink[1]{}%
\providecommand \@@endlink[0]{}%
\providecommand \url  [0]{\begingroup\@sanitize@url \@url }%
\providecommand \@url [1]{\endgroup\@href {#1}{\urlprefix }}%
\providecommand \urlprefix  [0]{URL }%
\providecommand \Eprint [0]{\href }%
\providecommand \doibase [0]{http://dx.doi.org/}%
\providecommand \selectlanguage [0]{\@gobble}%
\providecommand \bibinfo  [0]{\@secondoftwo}%
\providecommand \bibfield  [0]{\@secondoftwo}%
\providecommand \translation [1]{[#1]}%
\providecommand \BibitemOpen [0]{}%
\providecommand \bibitemStop [0]{}%
\providecommand \bibitemNoStop [0]{.\EOS\space}%
\providecommand \EOS [0]{\spacefactor3000\relax}%
\providecommand \BibitemShut  [1]{\csname bibitem#1\endcsname}%
\let\auto@bib@innerbib\@empty
\bibitem [{\citenamefont {Casida}(1995)}]{CasidaTDDFT1995}%
  \BibitemOpen
  \bibfield  {author} {\bibinfo {author} {\bibfnamefont {M.~E.}\ \bibnamefont
  {Casida}},\ }in\ \href {\doibase 10.1142/9789812830586_0005} {\emph {\bibinfo
  {booktitle} {Recent Advances in Density Functional Methods}}}\ (\bibinfo
  {publisher} {World Scientific},\ \bibinfo {year} {1995})\ Chap.~\bibinfo
  {chapter} {5}, pp.\ \bibinfo {pages} {155--192}\BibitemShut {NoStop}%
\bibitem [{\citenamefont {Kasha}(1950)}]{KashaRule1950}%
  \BibitemOpen
  \bibfield  {author} {\bibinfo {author} {\bibfnamefont {M.}~\bibnamefont
  {Kasha}},\ }\href {\doibase 10.1039/df9500900014} {\bibfield  {journal}
  {\bibinfo  {journal} {Discuss. Faraday Soc.}\ }\textbf {\bibinfo {volume}
  {9}},\ \bibinfo {pages} {14} (\bibinfo {year} {1950})}\BibitemShut {NoStop}%
\bibitem [{\citenamefont {Porezag}\ \emph {et~al.}(1995)\citenamefont
  {Porezag}, \citenamefont {Frauenheim}, \citenamefont {K\"ohler},
  \citenamefont {Seifert},\ and\ \citenamefont {Kaschner}}]{PorezagDFTB1995}%
  \BibitemOpen
  \bibfield  {author} {\bibinfo {author} {\bibfnamefont {D.}~\bibnamefont
  {Porezag}}, \bibinfo {author} {\bibfnamefont {T.}~\bibnamefont {Frauenheim}},
  \bibinfo {author} {\bibfnamefont {T.}~\bibnamefont {K\"ohler}}, \bibinfo
  {author} {\bibfnamefont {G.}~\bibnamefont {Seifert}}, \ and\ \bibinfo
  {author} {\bibfnamefont {R.}~\bibnamefont {Kaschner}},\ }\href {\doibase
  10.1103/PhysRevB.51.12947} {\bibfield  {journal} {\bibinfo  {journal} {Phys.
  Rev. B}\ }\textbf {\bibinfo {volume} {51}},\ \bibinfo {pages} {12947}
  (\bibinfo {year} {1995})}\BibitemShut {NoStop}%
\bibitem [{\citenamefont {Seifert}, \citenamefont {Porezag},\ and\
  \citenamefont {Frauenheim}(1996)}]{SeifertDFTB1996}%
  \BibitemOpen
  \bibfield  {author} {\bibinfo {author} {\bibfnamefont {G.}~\bibnamefont
  {Seifert}}, \bibinfo {author} {\bibfnamefont {D.}~\bibnamefont {Porezag}}, \
  and\ \bibinfo {author} {\bibfnamefont {T.}~\bibnamefont {Frauenheim}},\
  }\href {\doibase 10.1002/(SICI)1097-461X(1996)58:2<185::AID-QUA7>3.0.CO;2-U}
  {\bibfield  {journal} {\bibinfo  {journal} {Int. J. Quantum Chem.}\ }\textbf
  {\bibinfo {volume} {58}},\ \bibinfo {pages} {185} (\bibinfo {year}
  {1996})}\BibitemShut {NoStop}%
\bibitem [{\citenamefont {Niehaus}\ \emph {et~al.}(2001)\citenamefont
  {Niehaus}, \citenamefont {Suhai}, \citenamefont {Della~Sala}, \citenamefont
  {Lugli}, \citenamefont {Elstner}, \citenamefont {Seifert},\ and\
  \citenamefont {Frauenheim}}]{NiehausTDDFTB2001}%
  \BibitemOpen
  \bibfield  {author} {\bibinfo {author} {\bibfnamefont {T.~A.}\ \bibnamefont
  {Niehaus}}, \bibinfo {author} {\bibfnamefont {S.}~\bibnamefont {Suhai}},
  \bibinfo {author} {\bibfnamefont {F.}~\bibnamefont {Della~Sala}}, \bibinfo
  {author} {\bibfnamefont {P.}~\bibnamefont {Lugli}}, \bibinfo {author}
  {\bibfnamefont {M.}~\bibnamefont {Elstner}}, \bibinfo {author} {\bibfnamefont
  {G.}~\bibnamefont {Seifert}}, \ and\ \bibinfo {author} {\bibfnamefont
  {T.}~\bibnamefont {Frauenheim}},\ }\href {\doibase
  10.1103/PhysRevB.63.085108} {\bibfield  {journal} {\bibinfo  {journal} {Phys.
  Rev. B}\ }\textbf {\bibinfo {volume} {63}},\ \bibinfo {pages} {085108}
  (\bibinfo {year} {2001})}\BibitemShut {NoStop}%
\bibitem [{\citenamefont {Joswig}\ \emph {et~al.}(2003)\citenamefont {Joswig},
  \citenamefont {Seifert}, \citenamefont {Niehaus},\ and\ \citenamefont
  {Springborg}}]{tddftbapp1_doi:10.1021/jp026752s}%
  \BibitemOpen
  \bibfield  {author} {\bibinfo {author} {\bibfnamefont {J.-O.}\ \bibnamefont
  {Joswig}}, \bibinfo {author} {\bibfnamefont {G.}~\bibnamefont {Seifert}},
  \bibinfo {author} {\bibfnamefont {T.~A.}\ \bibnamefont {Niehaus}}, \ and\
  \bibinfo {author} {\bibfnamefont {M.}~\bibnamefont {Springborg}},\ }\href
  {\doibase 10.1021/jp026752s} {\bibfield  {journal} {\bibinfo  {journal} {J.
  Phys. Chem. B}\ }\textbf {\bibinfo {volume} {107}},\ \bibinfo {pages} {2897}
  (\bibinfo {year} {2003})}\BibitemShut {NoStop}%
\bibitem [{\citenamefont {Goswami}\ \emph {et~al.}(2006)\citenamefont
  {Goswami}, \citenamefont {Pal}, \citenamefont {Sarkar}, \citenamefont
  {Seifert},\ and\ \citenamefont {Springborg}}]{tddftbapp2_PhysRevB.73.205312}%
  \BibitemOpen
  \bibfield  {author} {\bibinfo {author} {\bibfnamefont {B.}~\bibnamefont
  {Goswami}}, \bibinfo {author} {\bibfnamefont {S.}~\bibnamefont {Pal}},
  \bibinfo {author} {\bibfnamefont {P.}~\bibnamefont {Sarkar}}, \bibinfo
  {author} {\bibfnamefont {G.}~\bibnamefont {Seifert}}, \ and\ \bibinfo
  {author} {\bibfnamefont {M.}~\bibnamefont {Springborg}},\ }\href {\doibase
  10.1103/PhysRevB.73.205312} {\bibfield  {journal} {\bibinfo  {journal} {Phys.
  Rev. B}\ }\textbf {\bibinfo {volume} {73}},\ \bibinfo {pages} {205312}
  (\bibinfo {year} {2006})}\BibitemShut {NoStop}%
\bibitem [{\citenamefont {Frenzel}, \citenamefont {Joswig},\ and\ \citenamefont
  {Seifert}(2007)}]{tddftbapp3_doi:10.1021/jp071125u}%
  \BibitemOpen
  \bibfield  {author} {\bibinfo {author} {\bibfnamefont {J.}~\bibnamefont
  {Frenzel}}, \bibinfo {author} {\bibfnamefont {J.-O.}\ \bibnamefont {Joswig}},
  \ and\ \bibinfo {author} {\bibfnamefont {G.}~\bibnamefont {Seifert}},\ }\href
  {\doibase 10.1021/jp071125u} {\bibfield  {journal} {\bibinfo  {journal} {J.
  Phys. Chem. C}\ }\textbf {\bibinfo {volume} {111}},\ \bibinfo {pages} {10761}
  (\bibinfo {year} {2007})}\BibitemShut {NoStop}%
\bibitem [{\citenamefont {Li}\ \emph {et~al.}(2007)\citenamefont {Li},
  \citenamefont {Zhang}, \citenamefont {Niehaus}, \citenamefont {Frauenheim},\
  and\ \citenamefont {Lee}}]{tddftbapp4_doi:10.1021/ct700041v}%
  \BibitemOpen
  \bibfield  {author} {\bibinfo {author} {\bibfnamefont {Q.~S.}\ \bibnamefont
  {Li}}, \bibinfo {author} {\bibfnamefont {R.~Q.}\ \bibnamefont {Zhang}},
  \bibinfo {author} {\bibfnamefont {T.~A.}\ \bibnamefont {Niehaus}}, \bibinfo
  {author} {\bibfnamefont {T.}~\bibnamefont {Frauenheim}}, \ and\ \bibinfo
  {author} {\bibfnamefont {S.~T.}\ \bibnamefont {Lee}},\ }\href {\doibase
  10.1021/ct700041v} {\bibfield  {journal} {\bibinfo  {journal} {J. Chem.
  Theory Comput.}\ }\textbf {\bibinfo {volume} {3}},\ \bibinfo {pages} {1518}
  (\bibinfo {year} {2007})}\BibitemShut {NoStop}%
\bibitem [{\citenamefont {Wang}\ \emph
  {et~al.}(2007{\natexlab{a}})\citenamefont {Wang}, \citenamefont {Zhang},
  \citenamefont {Niehaus},\ and\ \citenamefont
  {Frauenheim}}]{tddftbapp5_doi:10.1021/jp065704v}%
  \BibitemOpen
  \bibfield  {author} {\bibinfo {author} {\bibfnamefont {X.}~\bibnamefont
  {Wang}}, \bibinfo {author} {\bibfnamefont {R.~Q.}\ \bibnamefont {Zhang}},
  \bibinfo {author} {\bibfnamefont {T.~A.}\ \bibnamefont {Niehaus}}, \ and\
  \bibinfo {author} {\bibfnamefont {T.}~\bibnamefont {Frauenheim}},\ }\href
  {http://pubs.acs.org/doi/abs/10.1021/jp065704v} {\bibfield  {journal}
  {\bibinfo  {journal} {J. Phys. Chem. C}\ }\textbf {\bibinfo {volume} {111}},\
  \bibinfo {pages} {2394} (\bibinfo {year} {2007}{\natexlab{a}})}\BibitemShut
  {NoStop}%
\bibitem [{\citenamefont {Wang}\ \emph
  {et~al.}(2007{\natexlab{b}})\citenamefont {Wang}, \citenamefont {Zhang},
  \citenamefont {Lee}, \citenamefont {Niehaus},\ and\ \citenamefont
  {Frauenheim}}]{tddftbapp6_10.1063/1.2715101}%
  \BibitemOpen
  \bibfield  {author} {\bibinfo {author} {\bibfnamefont {X.}~\bibnamefont
  {Wang}}, \bibinfo {author} {\bibfnamefont {R.~Q.}\ \bibnamefont {Zhang}},
  \bibinfo {author} {\bibfnamefont {S.~T.}\ \bibnamefont {Lee}}, \bibinfo
  {author} {\bibfnamefont {T.~A.}\ \bibnamefont {Niehaus}}, \ and\ \bibinfo
  {author} {\bibfnamefont {T.}~\bibnamefont {Frauenheim}},\ }\href
  {http://scitation.aip.org/content/aip/journal/apl/90/12/10.1063/1.2715101}
  {\bibfield  {journal} {\bibinfo  {journal} {Appl. Phys. Lett.}\ }\textbf
  {\bibinfo {volume} {90}},\ \bibinfo {eid} {123116} (\bibinfo {year}
  {2007}{\natexlab{b}})}\BibitemShut {NoStop}%
\bibitem [{\citenamefont {Li}\ \emph {et~al.}(2008)\citenamefont {Li},
  \citenamefont {Zhang}, \citenamefont {Lee}, \citenamefont {Niehaus},\ and\
  \citenamefont {Frauenheim}}]{tddftbapp7_doi:10.1063/1.2940735}%
  \BibitemOpen
  \bibfield  {author} {\bibinfo {author} {\bibfnamefont {Q.~S.}\ \bibnamefont
  {Li}}, \bibinfo {author} {\bibfnamefont {R.~Q.}\ \bibnamefont {Zhang}},
  \bibinfo {author} {\bibfnamefont {S.~T.}\ \bibnamefont {Lee}}, \bibinfo
  {author} {\bibfnamefont {T.~A.}\ \bibnamefont {Niehaus}}, \ and\ \bibinfo
  {author} {\bibfnamefont {T.}~\bibnamefont {Frauenheim}},\ }\href
  {http://scitation.aip.org/content/aip/journal/jcp/128/24/10.1063/1.2940735}
  {\bibfield  {journal} {\bibinfo  {journal} {J. Chem. Phys.}\ }\textbf
  {\bibinfo {volume} {128}},\ \bibinfo {eid} {244714} (\bibinfo {year}
  {2008})}\BibitemShut {NoStop}%
\bibitem [{\citenamefont {Mitrić}\ \emph {et~al.}(2009)\citenamefont
  {Mitrić}, \citenamefont {Werner}, \citenamefont {Wohlgemuth}, \citenamefont
  {Seifert},\ and\ \citenamefont
  {Bonačić-Koutecký}}]{BonacicNonAdMDWithTDDFTB2009}%
  \BibitemOpen
  \bibfield  {author} {\bibinfo {author} {\bibfnamefont {R.}~\bibnamefont
  {Mitrić}}, \bibinfo {author} {\bibfnamefont {U.}~\bibnamefont {Werner}},
  \bibinfo {author} {\bibfnamefont {M.}~\bibnamefont {Wohlgemuth}}, \bibinfo
  {author} {\bibfnamefont {G.}~\bibnamefont {Seifert}}, \ and\ \bibinfo
  {author} {\bibfnamefont {V.}~\bibnamefont {Bonačić-Koutecký}},\ }\href
  {\doibase 10.1021/jp905600w} {\bibfield  {journal} {\bibinfo  {journal} {J.
  Phys. Chem. A}\ }\textbf {\bibinfo {volume} {113}},\ \bibinfo {pages} {12700}
  (\bibinfo {year} {2009})}\BibitemShut {NoStop}%
\bibitem [{\citenamefont {Zhang}\ \emph {et~al.}(2012)\citenamefont {Zhang},
  \citenamefont {De~Sarkar}, \citenamefont {Niehaus},\ and\ \citenamefont
  {Frauenheim}}]{tddftbapp9_PSSB:PSSB201100719}%
  \BibitemOpen
  \bibfield  {author} {\bibinfo {author} {\bibfnamefont {R.-Q.}\ \bibnamefont
  {Zhang}}, \bibinfo {author} {\bibfnamefont {A.}~\bibnamefont {De~Sarkar}},
  \bibinfo {author} {\bibfnamefont {T.~A.}\ \bibnamefont {Niehaus}}, \ and\
  \bibinfo {author} {\bibfnamefont {T.}~\bibnamefont {Frauenheim}},\ }\href
  {\doibase 10.1002/pssb.201100719} {\bibfield  {journal} {\bibinfo  {journal}
  {Phys. Status Solidi B}\ }\textbf {\bibinfo {volume} {249}},\ \bibinfo
  {pages} {401} (\bibinfo {year} {2012})}\BibitemShut {NoStop}%
\bibitem [{\citenamefont {Fan}\ \emph {et~al.}(2014)\citenamefont {Fan},
  \citenamefont {Li}, \citenamefont {Liu},\ and\ \citenamefont
  {He}}]{tddftbapp10_Fan201417}%
  \BibitemOpen
  \bibfield  {author} {\bibinfo {author} {\bibfnamefont {G.-H.}\ \bibnamefont
  {Fan}}, \bibinfo {author} {\bibfnamefont {X.}~\bibnamefont {Li}}, \bibinfo
  {author} {\bibfnamefont {J.-Y.}\ \bibnamefont {Liu}}, \ and\ \bibinfo
  {author} {\bibfnamefont {G.-Z.}\ \bibnamefont {He}},\ }\href {\doibase
  10.1016/j.comptc.2013.12.010} {\bibfield  {journal} {\bibinfo  {journal}
  {Comp. Theor. Chem.}\ }\textbf {\bibinfo {volume} {1030}},\ \bibinfo {pages}
  {17} (\bibinfo {year} {2014})}\BibitemShut {NoStop}%
\bibitem [{\citenamefont {Niehaus}(2009)}]{NiehausTDDFTBReview2009}%
  \BibitemOpen
  \bibfield  {author} {\bibinfo {author} {\bibfnamefont {T.~A.}\ \bibnamefont
  {Niehaus}},\ }\href {\doibase 10.1016/j.theochem.2009.04.034} {\bibfield
  {journal} {\bibinfo  {journal} {J. Mol. Struc.: THEOCHEM}\ }\textbf {\bibinfo
  {volume} {914}},\ \bibinfo {pages} {38} (\bibinfo {year} {2009})}\BibitemShut
  {NoStop}%
\bibitem [{\citenamefont {Dierksen}\ and\ \citenamefont
  {Grimme}(2004{\natexlab{a}})}]{DierksenVibronicDFT2004}%
  \BibitemOpen
  \bibfield  {author} {\bibinfo {author} {\bibfnamefont {M.}~\bibnamefont
  {Dierksen}}\ and\ \bibinfo {author} {\bibfnamefont {S.}~\bibnamefont
  {Grimme}},\ }\href {\doibase 10.1063/1.1642595} {\bibfield  {journal}
  {\bibinfo  {journal} {J. Chem. Phys.}\ }\textbf {\bibinfo {volume} {120}},\
  \bibinfo {pages} {3544} (\bibinfo {year} {2004}{\natexlab{a}})}\BibitemShut
  {NoStop}%
\bibitem [{\citenamefont {Dierksen}\ and\ \citenamefont
  {Grimme}(2004{\natexlab{b}})}]{DierksenVibronicEXX2004}%
  \BibitemOpen
  \bibfield  {author} {\bibinfo {author} {\bibfnamefont {M.}~\bibnamefont
  {Dierksen}}\ and\ \bibinfo {author} {\bibfnamefont {S.}~\bibnamefont
  {Grimme}},\ }\href {\doibase 10.1021/jp047289h} {\bibfield  {journal}
  {\bibinfo  {journal} {J. Phys. Chem. A}\ }\textbf {\bibinfo {volume} {108}},\
  \bibinfo {pages} {10225} (\bibinfo {year} {2004}{\natexlab{b}})}\BibitemShut
  {NoStop}%
\bibitem [{\citenamefont {Barone}\ \emph {et~al.}(2009)\citenamefont {Barone},
  \citenamefont {Bloino}, \citenamefont {Biczysko},\ and\ \citenamefont
  {Santoro}}]{BaroneFrankCondon2009}%
  \BibitemOpen
  \bibfield  {author} {\bibinfo {author} {\bibfnamefont {V.}~\bibnamefont
  {Barone}}, \bibinfo {author} {\bibfnamefont {J.}~\bibnamefont {Bloino}},
  \bibinfo {author} {\bibfnamefont {M.}~\bibnamefont {Biczysko}}, \ and\
  \bibinfo {author} {\bibfnamefont {F.}~\bibnamefont {Santoro}},\ }\href
  {\doibase 10.1021/ct8004744} {\bibfield  {journal} {\bibinfo  {journal} {J.
  Chem. Theory Comput.}\ }\textbf {\bibinfo {volume} {5}},\ \bibinfo {pages}
  {540} (\bibinfo {year} {2009})}\BibitemShut {NoStop}%
\bibitem [{\citenamefont {Bloino}\ \emph {et~al.}(2010)\citenamefont {Bloino},
  \citenamefont {Biczysko}, \citenamefont {Santoro},\ and\ \citenamefont
  {Barone}}]{BaroneFrankCondon2010}%
  \BibitemOpen
  \bibfield  {author} {\bibinfo {author} {\bibfnamefont {J.}~\bibnamefont
  {Bloino}}, \bibinfo {author} {\bibfnamefont {M.}~\bibnamefont {Biczysko}},
  \bibinfo {author} {\bibfnamefont {F.}~\bibnamefont {Santoro}}, \ and\
  \bibinfo {author} {\bibfnamefont {V.}~\bibnamefont {Barone}},\ }\href
  {\doibase 10.1021/ct9006772} {\bibfield  {journal} {\bibinfo  {journal} {J.
  Chem. Theory Comput.}\ }\textbf {\bibinfo {volume} {6}},\ \bibinfo {pages}
  {1256} (\bibinfo {year} {2010})}\BibitemShut {NoStop}%
\bibitem [{\citenamefont {Stendardo}\ \emph {et~al.}(2012)\citenamefont
  {Stendardo}, \citenamefont {Ferrer}, \citenamefont {Santoro},\ and\
  \citenamefont {Improta}}]{StendardoVibResDithiophene2012}%
  \BibitemOpen
  \bibfield  {author} {\bibinfo {author} {\bibfnamefont {E.}~\bibnamefont
  {Stendardo}}, \bibinfo {author} {\bibfnamefont {F.~A.}\ \bibnamefont
  {Ferrer}}, \bibinfo {author} {\bibfnamefont {F.}~\bibnamefont {Santoro}}, \
  and\ \bibinfo {author} {\bibfnamefont {R.}~\bibnamefont {Improta}},\ }\href
  {\doibase 10.1021/ct300664d} {\bibfield  {journal} {\bibinfo  {journal} {J.
  Chem. Theory Comput.}\ }\textbf {\bibinfo {volume} {8}},\ \bibinfo {pages}
  {4483} (\bibinfo {year} {2012})}\BibitemShut {NoStop}%
\bibitem [{\citenamefont {Muniz-Miranda}\ \emph {et~al.}(2015)\citenamefont
  {Muniz-Miranda}, \citenamefont {Pedone}, \citenamefont {Battistelli},
  \citenamefont {Montalti}, \citenamefont {Bloino},\ and\ \citenamefont
  {Barone}}]{BaroneVibResSpecComparison2015}%
  \BibitemOpen
  \bibfield  {author} {\bibinfo {author} {\bibfnamefont {F.}~\bibnamefont
  {Muniz-Miranda}}, \bibinfo {author} {\bibfnamefont {A.}~\bibnamefont
  {Pedone}}, \bibinfo {author} {\bibfnamefont {G.}~\bibnamefont {Battistelli}},
  \bibinfo {author} {\bibfnamefont {M.}~\bibnamefont {Montalti}}, \bibinfo
  {author} {\bibfnamefont {J.}~\bibnamefont {Bloino}}, \ and\ \bibinfo {author}
  {\bibfnamefont {V.}~\bibnamefont {Barone}},\ }\href {\doibase
  10.1021/acs.jctc.5b00750} {\bibfield  {journal} {\bibinfo  {journal} {J.
  Chem. Theory Comput.}\ }\textbf {\bibinfo {volume} {11}},\ \bibinfo {pages}
  {5371} (\bibinfo {year} {2015})}\BibitemShut {NoStop}%
\bibitem [{\citenamefont {Heringer}\ \emph {et~al.}(2007)\citenamefont
  {Heringer}, \citenamefont {Niehaus}, \citenamefont {Wanko},\ and\
  \citenamefont {Frauenheim}}]{NiehausTDDFTBForces2007}%
  \BibitemOpen
  \bibfield  {author} {\bibinfo {author} {\bibfnamefont {D.}~\bibnamefont
  {Heringer}}, \bibinfo {author} {\bibfnamefont {T.~A.}\ \bibnamefont
  {Niehaus}}, \bibinfo {author} {\bibfnamefont {M.}~\bibnamefont {Wanko}}, \
  and\ \bibinfo {author} {\bibfnamefont {T.}~\bibnamefont {Frauenheim}},\
  }\href {\doibase 10.1002/jcc.20697} {\bibfield  {journal} {\bibinfo
  {journal} {J. Comput. Chem.}\ }\textbf {\bibinfo {volume} {28}},\ \bibinfo
  {pages} {2589} (\bibinfo {year} {2007})}\BibitemShut {NoStop}%
\bibitem [{\citenamefont {Mulliken}(1955)}]{MullikenPopulationAnalysis1955}%
  \BibitemOpen
  \bibfield  {author} {\bibinfo {author} {\bibfnamefont {R.~S.}\ \bibnamefont
  {Mulliken}},\ }\href {\doibase 10.1063/1.1740588} {\bibfield  {journal}
  {\bibinfo  {journal} {J. Chem. Phys.}\ }\textbf {\bibinfo {volume} {23}},\
  \bibinfo {pages} {1833} (\bibinfo {year} {1955})}\BibitemShut {NoStop}%
\bibitem [{\citenamefont {Elstner}\ \emph {et~al.}(1998)\citenamefont
  {Elstner}, \citenamefont {Porezag}, \citenamefont {Jungnickel}, \citenamefont
  {Elsner}, \citenamefont {Haugk}, \citenamefont {Frauenheim}, \citenamefont
  {Suhai},\ and\ \citenamefont {Seifert}}]{SeifertSCCDFTB1998}%
  \BibitemOpen
  \bibfield  {author} {\bibinfo {author} {\bibfnamefont {M.}~\bibnamefont
  {Elstner}}, \bibinfo {author} {\bibfnamefont {D.}~\bibnamefont {Porezag}},
  \bibinfo {author} {\bibfnamefont {G.}~\bibnamefont {Jungnickel}}, \bibinfo
  {author} {\bibfnamefont {J.}~\bibnamefont {Elsner}}, \bibinfo {author}
  {\bibfnamefont {M.}~\bibnamefont {Haugk}}, \bibinfo {author} {\bibfnamefont
  {T.}~\bibnamefont {Frauenheim}}, \bibinfo {author} {\bibfnamefont
  {S.}~\bibnamefont {Suhai}}, \ and\ \bibinfo {author} {\bibfnamefont
  {G.}~\bibnamefont {Seifert}},\ }\href {\doibase 10.1103/PhysRevB.58.7260}
  {\bibfield  {journal} {\bibinfo  {journal} {Phys. Rev. B}\ }\textbf {\bibinfo
  {volume} {58}},\ \bibinfo {pages} {7260} (\bibinfo {year}
  {1998})}\BibitemShut {NoStop}%
\bibitem [{\citenamefont {Köhler}\ \emph {et~al.}(2001)\citenamefont
  {Köhler}, \citenamefont {Seifert}, \citenamefont {Gerstmann}, \citenamefont
  {Elstner}, \citenamefont {Overhof},\ and\ \citenamefont
  {Frauenheim}}]{KohlerSDFTB2001}%
  \BibitemOpen
  \bibfield  {author} {\bibinfo {author} {\bibfnamefont {C.}~\bibnamefont
  {Köhler}}, \bibinfo {author} {\bibfnamefont {G.}~\bibnamefont {Seifert}},
  \bibinfo {author} {\bibfnamefont {U.}~\bibnamefont {Gerstmann}}, \bibinfo
  {author} {\bibfnamefont {M.}~\bibnamefont {Elstner}}, \bibinfo {author}
  {\bibfnamefont {H.}~\bibnamefont {Overhof}}, \ and\ \bibinfo {author}
  {\bibfnamefont {T.}~\bibnamefont {Frauenheim}},\ }\href {\doibase
  10.1039/B105782K} {\bibfield  {journal} {\bibinfo  {journal} {Phys. Chem.
  Chem. Phys.}\ }\textbf {\bibinfo {volume} {3}},\ \bibinfo {pages} {5109}
  (\bibinfo {year} {2001})}\BibitemShut {NoStop}%
\bibitem [{\citenamefont {Furche}\ and\ \citenamefont
  {Ahlrichs}(2002)}]{FurcheTDDFTForces2002}%
  \BibitemOpen
  \bibfield  {author} {\bibinfo {author} {\bibfnamefont {F.}~\bibnamefont
  {Furche}}\ and\ \bibinfo {author} {\bibfnamefont {R.}~\bibnamefont
  {Ahlrichs}},\ }\href {\doibase 10.1063/1.1508368} {\bibfield  {journal}
  {\bibinfo  {journal} {J. Chem. Phys.}\ }\textbf {\bibinfo {volume} {117}},\
  \bibinfo {pages} {7433} (\bibinfo {year} {2002})}\BibitemShut {NoStop}%
\bibitem [{\citenamefont {Born}\ and\ \citenamefont
  {Oppenheimer}(1927)}]{BornOppenheimer1927}%
  \BibitemOpen
  \bibfield  {author} {\bibinfo {author} {\bibfnamefont {M.}~\bibnamefont
  {Born}}\ and\ \bibinfo {author} {\bibfnamefont {R.}~\bibnamefont
  {Oppenheimer}},\ }\href {\doibase 10.1002/andp.19273892002} {\bibfield
  {journal} {\bibinfo  {journal} {Annalen der Physik}\ }\textbf {\bibinfo
  {volume} {389}},\ \bibinfo {pages} {457} (\bibinfo {year}
  {1927})}\BibitemShut {NoStop}%
\bibitem [{\citenamefont {Franck}\ and\ \citenamefont
  {Dymond}(1926)}]{FranckFCF1926}%
  \BibitemOpen
  \bibfield  {author} {\bibinfo {author} {\bibfnamefont {J.}~\bibnamefont
  {Franck}}\ and\ \bibinfo {author} {\bibfnamefont {E.~G.}\ \bibnamefont
  {Dymond}},\ }\href {\doibase 10.1039/TF9262100536} {\bibfield  {journal}
  {\bibinfo  {journal} {Trans. Faraday Soc.}\ }\textbf {\bibinfo {volume}
  {21}},\ \bibinfo {pages} {536} (\bibinfo {year} {1926})}\BibitemShut
  {NoStop}%
\bibitem [{\citenamefont {Condon}(1926)}]{CondonFCF1926}%
  \BibitemOpen
  \bibfield  {author} {\bibinfo {author} {\bibfnamefont {E.}~\bibnamefont
  {Condon}},\ }\href {\doibase 10.1103/PhysRev.28.1182} {\bibfield  {journal}
  {\bibinfo  {journal} {Phys. Rev.}\ }\textbf {\bibinfo {volume} {28}},\
  \bibinfo {pages} {1182} (\bibinfo {year} {1926})}\BibitemShut {NoStop}%
\bibitem [{\citenamefont {Condon}(1928)}]{CondonFCF1928}%
  \BibitemOpen
  \bibfield  {author} {\bibinfo {author} {\bibfnamefont {E.~U.}\ \bibnamefont
  {Condon}},\ }\href {\doibase 10.1103/PhysRev.32.858} {\bibfield  {journal}
  {\bibinfo  {journal} {Phys. Rev.}\ }\textbf {\bibinfo {volume} {32}},\
  \bibinfo {pages} {858} (\bibinfo {year} {1928})}\BibitemShut {NoStop}%
\bibitem [{\citenamefont {Herzberg}\ and\ \citenamefont
  {Teller}(1933)}]{HerzbergTeller1933}%
  \BibitemOpen
  \bibfield  {author} {\bibinfo {author} {\bibfnamefont {G.}~\bibnamefont
  {Herzberg}}\ and\ \bibinfo {author} {\bibfnamefont {E.}~\bibnamefont
  {Teller}},\ }\href@noop {} {\bibfield  {journal} {\bibinfo  {journal} {Z.
  Phys. Chem. Abt. B}\ }\textbf {\bibinfo {volume} {21}},\ \bibinfo {pages}
  {410} (\bibinfo {year} {1933})}\BibitemShut {NoStop}%
\bibitem [{\citenamefont {Duschinsky}(1937)}]{Duschinsky1937}%
  \BibitemOpen
  \bibfield  {author} {\bibinfo {author} {\bibfnamefont {F.}~\bibnamefont
  {Duschinsky}},\ }\href@noop {} {\bibfield  {journal} {\bibinfo  {journal}
  {Acta Physicochim. URSS}\ }\textbf {\bibinfo {volume} {7}},\ \bibinfo {pages}
  {551} (\bibinfo {year} {1937})}\BibitemShut {NoStop}%
\bibitem [{\citenamefont {Ruhoff}(1994)}]{RuhoffRecursiveFCF1994}%
  \BibitemOpen
  \bibfield  {author} {\bibinfo {author} {\bibfnamefont {P.~T.}\ \bibnamefont
  {Ruhoff}},\ }\href {\doibase 10.1016/0301-0104(94)00173-1} {\bibfield
  {journal} {\bibinfo  {journal} {Chem. Phys.}\ }\textbf {\bibinfo {volume}
  {186}},\ \bibinfo {pages} {355} (\bibinfo {year} {1994})}\BibitemShut
  {NoStop}%
\bibitem [{\citenamefont {Ruhoff}\ and\ \citenamefont
  {Ratner}(2000)}]{RuhoffRatnerFranckCondonAlgs2000}%
  \BibitemOpen
  \bibfield  {author} {\bibinfo {author} {\bibfnamefont {P.~T.}\ \bibnamefont
  {Ruhoff}}\ and\ \bibinfo {author} {\bibfnamefont {M.~A.}\ \bibnamefont
  {Ratner}},\ }\href {\doibase
  10.1002/(SICI)1097-461X(2000)77:1<383::AID-QUA38>3.0.CO;2-0} {\bibfield
  {journal} {\bibinfo  {journal} {Int. J. Quantum Chem.}\ }\textbf {\bibinfo
  {volume} {77}},\ \bibinfo {pages} {383} (\bibinfo {year} {2000})}\BibitemShut
  {NoStop}%
\bibitem [{\citenamefont {Macak}, \citenamefont {Luo},\ and\ \citenamefont
  {{\AA}gren}(2000)}]{MacakLinearCouplingModel2000}%
  \BibitemOpen
  \bibfield  {author} {\bibinfo {author} {\bibfnamefont {P.}~\bibnamefont
  {Macak}}, \bibinfo {author} {\bibfnamefont {Y.}~\bibnamefont {Luo}}, \ and\
  \bibinfo {author} {\bibfnamefont {H.}~\bibnamefont {{\AA}gren}},\ }\href
  {\doibase 10.1016/s0009-2614(00)01096-4} {\bibfield  {journal} {\bibinfo
  {journal} {Chem. Phys. Lett.}\ }\textbf {\bibinfo {volume} {330}},\ \bibinfo
  {pages} {447} (\bibinfo {year} {2000})}\BibitemShut {NoStop}%
\bibitem [{\citenamefont {Perdew}, \citenamefont {Burke},\ and\ \citenamefont
  {Ernzerhof}(1996)}]{PerdewBurkeErnzerhofPBEXCFunc1996}%
  \BibitemOpen
  \bibfield  {author} {\bibinfo {author} {\bibfnamefont {J.~P.}\ \bibnamefont
  {Perdew}}, \bibinfo {author} {\bibfnamefont {K.}~\bibnamefont {Burke}}, \
  and\ \bibinfo {author} {\bibfnamefont {M.}~\bibnamefont {Ernzerhof}},\ }\href
  {\doibase 10.1103/PhysRevLett.77.3865} {\bibfield  {journal} {\bibinfo
  {journal} {Phys. Rev. Lett.}\ }\textbf {\bibinfo {volume} {77}},\ \bibinfo
  {pages} {3865} (\bibinfo {year} {1996})}\BibitemShut {NoStop}%
\bibitem [{\citenamefont {Gaus}, \citenamefont {Cui},\ and\ \citenamefont
  {Elstner}(2011)}]{ElstnerDFTB32011}%
  \BibitemOpen
  \bibfield  {author} {\bibinfo {author} {\bibfnamefont {M.}~\bibnamefont
  {Gaus}}, \bibinfo {author} {\bibfnamefont {Q.}~\bibnamefont {Cui}}, \ and\
  \bibinfo {author} {\bibfnamefont {M.}~\bibnamefont {Elstner}},\ }\href
  {\doibase 10.1021/ct100684s} {\bibfield  {journal} {\bibinfo  {journal} {J.
  Chem. Theory Comput.}\ }\textbf {\bibinfo {volume} {7}},\ \bibinfo {pages}
  {931} (\bibinfo {year} {2011})}\BibitemShut {NoStop}%
\bibitem [{\citenamefont {Gaus}, \citenamefont {Goez},\ and\ \citenamefont
  {Elstner}(2013)}]{Elstner3obParameters2013}%
  \BibitemOpen
  \bibfield  {author} {\bibinfo {author} {\bibfnamefont {M.}~\bibnamefont
  {Gaus}}, \bibinfo {author} {\bibfnamefont {A.}~\bibnamefont {Goez}}, \ and\
  \bibinfo {author} {\bibfnamefont {M.}~\bibnamefont {Elstner}},\ }\href
  {\doibase 10.1021/ct300849w} {\bibfield  {journal} {\bibinfo  {journal} {J.
  Chem. Theory Comput.}\ }\textbf {\bibinfo {volume} {9}},\ \bibinfo {pages}
  {338} (\bibinfo {year} {2013})}\BibitemShut {NoStop}%
\bibitem [{\citenamefont {Nishimoto}(2015)}]{NishimotoTDDFTB32015}%
  \BibitemOpen
  \bibfield  {author} {\bibinfo {author} {\bibfnamefont {Y.}~\bibnamefont
  {Nishimoto}},\ }\href {\doibase 10.1063/1.4929926} {\bibfield  {journal}
  {\bibinfo  {journal} {J. Chem. Phys.}\ }\textbf {\bibinfo {volume} {143}},\
  \bibinfo {pages} {094108} (\bibinfo {year} {2015})}\BibitemShut {NoStop}%
\bibitem [{\citenamefont {Ferguson}, \citenamefont {Reeves},\ and\
  \citenamefont {Schneider}(1957)}]{ExpAnthracenePyrene}%
  \BibitemOpen
  \bibfield  {author} {\bibinfo {author} {\bibfnamefont {J.}~\bibnamefont
  {Ferguson}}, \bibinfo {author} {\bibfnamefont {L.~W.}\ \bibnamefont
  {Reeves}}, \ and\ \bibinfo {author} {\bibfnamefont {W.~G.}\ \bibnamefont
  {Schneider}},\ }\href {\doibase 10.1139/v57-152} {\bibfield  {journal}
  {\bibinfo  {journal} {Can. J. Chem.}\ }\textbf {\bibinfo {volume} {35}},\
  \bibinfo {pages} {1117} (\bibinfo {year} {1957})}\BibitemShut {NoStop}%
\bibitem [{\citenamefont {Banasiewicz}, \citenamefont {Deperasi\'{n}ska},\ and\
  \citenamefont {Kozankiewicz}(2003)}]{ExpPentacene}%
  \BibitemOpen
  \bibfield  {author} {\bibinfo {author} {\bibfnamefont {M.}~\bibnamefont
  {Banasiewicz}}, \bibinfo {author} {\bibfnamefont {I.}~\bibnamefont
  {Deperasi\'{n}ska}}, \ and\ \bibinfo {author} {\bibfnamefont
  {B.}~\bibnamefont {Kozankiewicz}},\ }\href {\doibase 10.1021/jp021527w}
  {\bibfield  {journal} {\bibinfo  {journal} {J. Phys. Chem. A}\ }\textbf
  {\bibinfo {volume} {107}},\ \bibinfo {pages} {662} (\bibinfo {year}
  {2003})}\BibitemShut {NoStop}%
\bibitem [{\citenamefont {Karabunarliev}\ \emph {et~al.}(1994)\citenamefont
  {Karabunarliev}, \citenamefont {Gherghel}, \citenamefont {Koch},\ and\
  \citenamefont {Baumgarten}}]{ExpPentarylene}%
  \BibitemOpen
  \bibfield  {author} {\bibinfo {author} {\bibfnamefont {S.}~\bibnamefont
  {Karabunarliev}}, \bibinfo {author} {\bibfnamefont {L.}~\bibnamefont
  {Gherghel}}, \bibinfo {author} {\bibfnamefont {K.-H.}\ \bibnamefont {Koch}},
  \ and\ \bibinfo {author} {\bibfnamefont {M.}~\bibnamefont {Baumgarten}},\
  }\href {\doibase 10.1016/0301-0104(94)80007-3} {\bibfield  {journal}
  {\bibinfo  {journal} {Chem. Phys.}\ }\textbf {\bibinfo {volume} {189}},\
  \bibinfo {pages} {53} (\bibinfo {year} {1994})}\BibitemShut {NoStop}%
\bibitem [{\citenamefont {Leopold}, \citenamefont {Vaida},\ and\ \citenamefont
  {Granville}(1984)}]{ExpOctatetraene}%
  \BibitemOpen
  \bibfield  {author} {\bibinfo {author} {\bibfnamefont {D.~G.}\ \bibnamefont
  {Leopold}}, \bibinfo {author} {\bibfnamefont {V.}~\bibnamefont {Vaida}}, \
  and\ \bibinfo {author} {\bibfnamefont {M.~F.}\ \bibnamefont {Granville}},\
  }\href {\doibase 10.1063/1.447452} {\bibfield  {journal} {\bibinfo  {journal}
  {J. Chem. Phys.}\ }\textbf {\bibinfo {volume} {81}},\ \bibinfo {pages} {4210}
  (\bibinfo {year} {1984})}\BibitemShut {NoStop}%
\bibitem [{\citenamefont {Dyck}\ and\ \citenamefont
  {McClure}(1962)}]{ExpStilbene}%
  \BibitemOpen
  \bibfield  {author} {\bibinfo {author} {\bibfnamefont {R.~H.}\ \bibnamefont
  {Dyck}}\ and\ \bibinfo {author} {\bibfnamefont {D.~S.}\ \bibnamefont
  {McClure}},\ }\href {\doibase 10.1063/1.1732885} {\bibfield  {journal}
  {\bibinfo  {journal} {J. Chem. Phys.}\ }\textbf {\bibinfo {volume} {36}},\
  \bibinfo {pages} {2326} (\bibinfo {year} {1962})}\BibitemShut {NoStop}%
\bibitem [{\citenamefont {Hoffmann}\ \emph {et~al.}(2006)\citenamefont
  {Hoffmann}, \citenamefont {Marquardt}, \citenamefont {Gemechu},\ and\
  \citenamefont {Baumgärtel}}]{AnisoleExp}%
  \BibitemOpen
  \bibfield  {author} {\bibinfo {author} {\bibfnamefont {L.~J.~H.}\
  \bibnamefont {Hoffmann}}, \bibinfo {author} {\bibfnamefont {S.}~\bibnamefont
  {Marquardt}}, \bibinfo {author} {\bibfnamefont {A.~S.}\ \bibnamefont
  {Gemechu}}, \ and\ \bibinfo {author} {\bibfnamefont {H.}~\bibnamefont
  {Baumgärtel}},\ }\href {\doibase 10.1039/b600438p} {\bibfield  {journal}
  {\bibinfo  {journal} {Phys. Chem. Chem. Phys.}\ }\textbf {\bibinfo {volume}
  {8}},\ \bibinfo {pages} {2360} (\bibinfo {year} {2006})}\BibitemShut
  {NoStop}%
\bibitem [{\citenamefont {Takayanagi}, \citenamefont {Gejo},\ and\
  \citenamefont {Hanazaki}(1994)}]{ExpDithiophene2}%
  \BibitemOpen
  \bibfield  {author} {\bibinfo {author} {\bibfnamefont {M.}~\bibnamefont
  {Takayanagi}}, \bibinfo {author} {\bibfnamefont {T.}~\bibnamefont {Gejo}}, \
  and\ \bibinfo {author} {\bibfnamefont {I.}~\bibnamefont {Hanazaki}},\ }\href
  {\doibase 10.1021/j100100a014} {\bibfield  {journal} {\bibinfo  {journal}
  {The Journal of Physical Chemistry}\ }\textbf {\bibinfo {volume} {98}},\
  \bibinfo {pages} {12893} (\bibinfo {year} {1994})}\BibitemShut {NoStop}%
\bibitem [{\citenamefont {Pocius}\ and\ \citenamefont
  {Yardley}(1973)}]{ExpTriazoline}%
  \BibitemOpen
  \bibfield  {author} {\bibinfo {author} {\bibfnamefont {A.~V.}\ \bibnamefont
  {Pocius}}\ and\ \bibinfo {author} {\bibfnamefont {J.~T.}\ \bibnamefont
  {Yardley}},\ }\href {\doibase 10.1021/ja00784a013} {\bibfield  {journal}
  {\bibinfo  {journal} {J. Am. Chem. Soc.}\ }\textbf {\bibinfo {volume} {95}},\
  \bibinfo {pages} {721} (\bibinfo {year} {1973})}\BibitemShut {NoStop}%
\bibitem [{\citenamefont {Iglesias-Groth}\ \emph {et~al.}(2010)\citenamefont
  {Iglesias-Groth}, \citenamefont {Manchado}, \citenamefont {Rebolo},
  \citenamefont {González~Hernández}, \citenamefont {García-Hernández},\
  and\ \citenamefont {Lambert}}]{AstroAnthracene}%
  \BibitemOpen
  \bibfield  {author} {\bibinfo {author} {\bibfnamefont {S.}~\bibnamefont
  {Iglesias-Groth}}, \bibinfo {author} {\bibfnamefont {A.}~\bibnamefont
  {Manchado}}, \bibinfo {author} {\bibfnamefont {R.}~\bibnamefont {Rebolo}},
  \bibinfo {author} {\bibfnamefont {J.~I.}\ \bibnamefont
  {González~Hernández}}, \bibinfo {author} {\bibfnamefont {D.~A.}\
  \bibnamefont {García-Hernández}}, \ and\ \bibinfo {author} {\bibfnamefont
  {D.~L.}\ \bibnamefont {Lambert}},\ }\href
  {http://mnras.oxfordjournals.org/content/407/4/2157.abstract} {\bibfield
  {journal} {\bibinfo  {journal} {Mon. Not. R. Astron. Soc.}\ }\textbf
  {\bibinfo {volume} {407}},\ \bibinfo {pages} {2157} (\bibinfo {year}
  {2010})}\BibitemShut {NoStop}%
\bibitem [{\citenamefont {Platt}(1949)}]{PlattNomenclature1949}%
  \BibitemOpen
  \bibfield  {author} {\bibinfo {author} {\bibfnamefont {J.~R.}\ \bibnamefont
  {Platt}},\ }\href {\doibase 10.1063/1.1747293} {\bibfield  {journal}
  {\bibinfo  {journal} {J. Chem. Phys.}\ }\textbf {\bibinfo {volume} {17}},\
  \bibinfo {pages} {484} (\bibinfo {year} {1949})}\BibitemShut {NoStop}%
\bibitem [{\citenamefont {Grimme}\ and\ \citenamefont
  {Parac}(2003)}]{GrimmeParacAcenesTDDFT2003}%
  \BibitemOpen
  \bibfield  {author} {\bibinfo {author} {\bibfnamefont {S.}~\bibnamefont
  {Grimme}}\ and\ \bibinfo {author} {\bibfnamefont {M.}~\bibnamefont {Parac}},\
  }\href {\doibase 10.1002/cphc.200390047} {\bibfield  {journal} {\bibinfo
  {journal} {{ChemPhysChem}}\ }\textbf {\bibinfo {volume} {4}},\ \bibinfo
  {pages} {292} (\bibinfo {year} {2003})}\BibitemShut {NoStop}%
\bibitem [{\citenamefont {Parac}\ and\ \citenamefont
  {Grimme}(2003)}]{ParacGrimmePAHWithTDDFT2003}%
  \BibitemOpen
  \bibfield  {author} {\bibinfo {author} {\bibfnamefont {M.}~\bibnamefont
  {Parac}}\ and\ \bibinfo {author} {\bibfnamefont {S.}~\bibnamefont {Grimme}},\
  }\href {\doibase 10.1016/s0301-0104(03)00250-7} {\bibfield  {journal}
  {\bibinfo  {journal} {Chem. Phys.}\ }\textbf {\bibinfo {volume} {292}},\
  \bibinfo {pages} {11} (\bibinfo {year} {2003})}\BibitemShut {NoStop}%
\bibitem [{\citenamefont {Peach}\ \emph {et~al.}(2008)\citenamefont {Peach},
  \citenamefont {Benfield}, \citenamefont {Helgaker},\ and\ \citenamefont
  {Tozer}}]{PeachTDDFTEval2008}%
  \BibitemOpen
  \bibfield  {author} {\bibinfo {author} {\bibfnamefont {M.~J.~G.}\
  \bibnamefont {Peach}}, \bibinfo {author} {\bibfnamefont {P.}~\bibnamefont
  {Benfield}}, \bibinfo {author} {\bibfnamefont {T.}~\bibnamefont {Helgaker}},
  \ and\ \bibinfo {author} {\bibfnamefont {D.~J.}\ \bibnamefont {Tozer}},\
  }\href {\doibase 10.1063/1.2831900} {\bibfield  {journal} {\bibinfo
  {journal} {J. Chem. Phys.}\ }\textbf {\bibinfo {volume} {128}},\ \bibinfo
  {pages} {044118} (\bibinfo {year} {2008})}\BibitemShut {NoStop}%
\bibitem [{\citenamefont {Geigle}, \citenamefont {Wolf},\ and\ \citenamefont
  {Hohlneicher}(1997)}]{GeiglePyreneLb1997}%
  \BibitemOpen
  \bibfield  {author} {\bibinfo {author} {\bibfnamefont {K.~P.}\ \bibnamefont
  {Geigle}}, \bibinfo {author} {\bibfnamefont {J.}~\bibnamefont {Wolf}}, \ and\
  \bibinfo {author} {\bibfnamefont {G.}~\bibnamefont {Hohlneicher}},\ }\href
  {\doibase 10.1016/s1010-6030(96)04607-2} {\bibfield  {journal} {\bibinfo
  {journal} {J. Photochem. Photobiol., A}\ }\textbf {\bibinfo {volume} {105}},\
  \bibinfo {pages} {183} (\bibinfo {year} {1997})}\BibitemShut {NoStop}%
\bibitem [{\citenamefont {Petek}\ \emph {et~al.}(1993)\citenamefont {Petek},
  \citenamefont {Bell}, \citenamefont {Choi}, \citenamefont {Yoshihara},
  \citenamefont {Tounge},\ and\ \citenamefont
  {Christensen}}]{Exp2Octatetraene}%
  \BibitemOpen
  \bibfield  {author} {\bibinfo {author} {\bibfnamefont {H.}~\bibnamefont
  {Petek}}, \bibinfo {author} {\bibfnamefont {A.~J.}\ \bibnamefont {Bell}},
  \bibinfo {author} {\bibfnamefont {Y.~S.}\ \bibnamefont {Choi}}, \bibinfo
  {author} {\bibfnamefont {K.}~\bibnamefont {Yoshihara}}, \bibinfo {author}
  {\bibfnamefont {B.~A.}\ \bibnamefont {Tounge}}, \ and\ \bibinfo {author}
  {\bibfnamefont {R.~L.}\ \bibnamefont {Christensen}},\ }\href {\doibase
  10.1063/1.464056} {\bibfield  {journal} {\bibinfo  {journal} {J. Chem.
  Phys.}\ }\textbf {\bibinfo {volume} {98}},\ \bibinfo {pages} {3777} (\bibinfo
  {year} {1993})}\BibitemShut {NoStop}%
\bibitem [{\citenamefont {Chai}\ and\ \citenamefont
  {Head-Gordon}(2008)}]{ChaiOmegaB97X2008}%
  \BibitemOpen
  \bibfield  {author} {\bibinfo {author} {\bibfnamefont {J.-D.}\ \bibnamefont
  {Chai}}\ and\ \bibinfo {author} {\bibfnamefont {M.}~\bibnamefont
  {Head-Gordon}},\ }\href {\doibase 10.1063/1.2834918} {\bibfield  {journal}
  {\bibinfo  {journal} {J. Chem. Phys.}\ }\textbf {\bibinfo {volume} {128}},\
  \bibinfo {pages} {084106} (\bibinfo {year} {2008})}\BibitemShut {NoStop}%
\bibitem [{\citenamefont {Birnbaum}\ and\ \citenamefont
  {Kohler}(1991)}]{ExpDithiophene}%
  \BibitemOpen
  \bibfield  {author} {\bibinfo {author} {\bibfnamefont {D.}~\bibnamefont
  {Birnbaum}}\ and\ \bibinfo {author} {\bibfnamefont {B.~E.}\ \bibnamefont
  {Kohler}},\ }\href {\doibase 10.1063/1.461721} {\bibfield  {journal}
  {\bibinfo  {journal} {J. Chem. Phys.}\ }\textbf {\bibinfo {volume} {95}},\
  \bibinfo {pages} {4783} (\bibinfo {year} {1991})}\BibitemShut {NoStop}%
\bibitem [{\citenamefont {Domínguez}\ \emph {et~al.}(2013)\citenamefont
  {Domínguez}, \citenamefont {Aradi}, \citenamefont {Frauenheim},
  \citenamefont {Lutsker},\ and\ \citenamefont
  {Niehaus}}]{NiehausTDDFTBOnsiteAndFracOcc2013}%
  \BibitemOpen
  \bibfield  {author} {\bibinfo {author} {\bibfnamefont {A.}~\bibnamefont
  {Domínguez}}, \bibinfo {author} {\bibfnamefont {B.}~\bibnamefont {Aradi}},
  \bibinfo {author} {\bibfnamefont {T.}~\bibnamefont {Frauenheim}}, \bibinfo
  {author} {\bibfnamefont {V.}~\bibnamefont {Lutsker}}, \ and\ \bibinfo
  {author} {\bibfnamefont {T.~A.}\ \bibnamefont {Niehaus}},\ }\href {\doibase
  10.1021/ct400123t} {\bibfield  {journal} {\bibinfo  {journal} {J. Chem.
  Theory Comput.}\ }\textbf {\bibinfo {volume} {9}},\ \bibinfo {pages} {4901}
  (\bibinfo {year} {2013})}\BibitemShut {NoStop}%
\bibitem [{\citenamefont {Gaus}\ \emph {et~al.}(2014)\citenamefont {Gaus},
  \citenamefont {Lu}, \citenamefont {Elstner},\ and\ \citenamefont
  {Cui}}]{Elstner3obSPParameters2014}%
  \BibitemOpen
  \bibfield  {author} {\bibinfo {author} {\bibfnamefont {M.}~\bibnamefont
  {Gaus}}, \bibinfo {author} {\bibfnamefont {X.}~\bibnamefont {Lu}}, \bibinfo
  {author} {\bibfnamefont {M.}~\bibnamefont {Elstner}}, \ and\ \bibinfo
  {author} {\bibfnamefont {Q.}~\bibnamefont {Cui}},\ }\href {\doibase
  10.1021/ct401002w} {\bibfield  {journal} {\bibinfo  {journal} {J. Chem.
  Theory Comput.}\ }\textbf {\bibinfo {volume} {10}},\ \bibinfo {pages} {1518}
  (\bibinfo {year} {2014})}\BibitemShut {NoStop}%
\bibitem [{\citenamefont {Niehaus}\ and\ \citenamefont
  {Sala}(2011)}]{NiehausLCDFTB2011}%
  \BibitemOpen
  \bibfield  {author} {\bibinfo {author} {\bibfnamefont {T.~A.}\ \bibnamefont
  {Niehaus}}\ and\ \bibinfo {author} {\bibfnamefont {F.~D.}\ \bibnamefont
  {Sala}},\ }\href {\doibase 10.1002/pssb.201100694} {\bibfield  {journal}
  {\bibinfo  {journal} {Phys. Status Solidi B}\ }\textbf {\bibinfo {volume}
  {249}},\ \bibinfo {pages} {237} (\bibinfo {year} {2011})}\BibitemShut
  {NoStop}%
\bibitem [{\citenamefont {Lutsker}, \citenamefont {Aradi},\ and\ \citenamefont
  {Niehaus}(2015)}]{LutskerLCDFTB2015}%
  \BibitemOpen
  \bibfield  {author} {\bibinfo {author} {\bibfnamefont {V.}~\bibnamefont
  {Lutsker}}, \bibinfo {author} {\bibfnamefont {B.}~\bibnamefont {Aradi}}, \
  and\ \bibinfo {author} {\bibfnamefont {T.~A.}\ \bibnamefont {Niehaus}},\
  }\href {\doibase 10.1063/1.4935095} {\bibfield  {journal} {\bibinfo
  {journal} {J. Chem. Phys.}\ }\textbf {\bibinfo {volume} {143}},\ \bibinfo
  {pages} {184107} (\bibinfo {year} {2015})}\BibitemShut {NoStop}%
\bibitem [{\citenamefont {Humeniuk}\ and\ \citenamefont
  {Mitri{\'{c}}}(2015)}]{HumeniukLCTDDFTB2015}%
  \BibitemOpen
  \bibfield  {author} {\bibinfo {author} {\bibfnamefont {A.}~\bibnamefont
  {Humeniuk}}\ and\ \bibinfo {author} {\bibfnamefont {R.}~\bibnamefont
  {Mitri{\'{c}}}},\ }\href {\doibase 10.1063/1.4931179} {\bibfield  {journal}
  {\bibinfo  {journal} {J. Chem. Phys.}\ }\textbf {\bibinfo {volume} {143}},\
  \bibinfo {pages} {134120} (\bibinfo {year} {2015})}\BibitemShut {NoStop}%
\end{thebibliography}%

\end{document}